\documentclass[useAMS,usenatbib,usegraphicx]{mn2e}

\pdfoutput=1
\usepackage{amsmath,amssymb}
\usepackage{bm}
\usepackage[usenames]{color}
\usepackage[caption=false]{subfig}
\usepackage{rotating}
\newcommand{\sgn}{\operatorname{sgn}}
\newcommand{\major}{black}
\newcommand{\minor}{black}

\title{Magnetic Turbulence and Thermodynamics \\ in the Inner Region of Protoplanetary Discs}

\author[S. Hirose]{Shigenobu Hirose$^{1}$\thanks{E-mail:hirose.shigenobu@gmail.com}\\
$^{1}$Department of Mathematical Science and Advanced Technology, Japan Agency for Marine-Earth Science and Technology,\\
  Yokohama, Kanagawa 236-0001, Japan}
\begin{document}



\maketitle

\label{firstpage}

\begin{abstract}
  Using radiation magnetohydrodynamics simulations with realistic opacities and equation of state, {\color{\minor}and zero net magnetic flux,}
  we have explored thermodynamics in the inner part of protoplanetary discs where magnetic turbulence is expected.
  The thermal equilibrium curve consists of the upper,
  lower, and middle branches. The upper (lower) branch corresponds to hot (cool) and optically {\color{\minor}very (moderately) thick} discs,
  respectively, while the middle branch is characterized by convective energy transport near the midplane.
  Convection is also {\color{\minor}the} major energy transport {\color{\minor} process} near the low surface density end of the upper branch.
  There, convective motion is fast {\color{\minor}with}
  Mach number{\color{\minor}s reaching $\gtrsim 0.01$}, {\color{\minor}and} enhances both magnetic turbulence and cooling,
  raising the ratio of vertically-integrated shear stress to vertically-integrated pressure by a factor of several.
  {\color{\minor}This convectively enhanced ratio} seems a robust feature in accretion discs {\color{\minor}having an} ionization transition.
  We have also examined causes of the S-shape{\color{\minor}d thermal equilibrium curve, as well as the thermal stability of
  the equilibrium solutions}.
  Finally, we compared our results with the disc instability models {\color{\minor}used to explain} FU Ori outbursts.
  {\color{\minor}Although the thermal equilibrium curve in our results also exhibits bistability,
    the surface density contrast across the bistability is an order of magnitude smaller, 
    and the stress-to-pressure ratios in both upper and lower branches are two orders of magnitude greater,
    than those favored in the disc instability models.}
  {\color{\minor}It therefore appears likely that} FU Ori outbursts {\color{\minor}are not due solely to a} thermal-viscous limit cycle
  {\color{\minor}resulting from} accretion driven by {\color{\minor}local} magnetic turbulence.
\end{abstract}


\section{Introduction}

Protoplanetary discs are accretion discs of {\color{\minor}mass accretion rates
  ranging from $2\times10^{-12}$ to $5\times10^{-8} M_\odot\text{yr}^{-1}$ and above \citep[e.g.][]{Herczeg08}}.
A most plausible mechanism for driving accretion discs is magnetic turbulence caused by magneto-rotational instability (MRI)
\citep[a comprehensive review by][and references are therein]{Balbus98}.
However, the ionization fraction of gas is generally low for most of radii in protoplanetary discs
since thermal ionization doesn't work due to low temperatures.
Therefore, non-ideal magnetohydrodynamics (MHD) effects 
are likely to suppress MRI there \citep[][recent works by, for example, \citet{Bai11,Wardle11,Kunz13}]{Sano99,Sano02,Sano03} although
non-thermal ionization by cosmic rays and the stellar X-rays may revive MRI in some limited places {\color{\minor}\citep{Gammie96,Sano00,Ilgner06,Turner08}}.
Pure hydrodynamic instabilities are also considered for such MRI-dead zones \citep{Nelson13,Klahr14}.

On the other hand,
MRI must play an important role on driving the mass accretion in the inner region 
where thermal ionization is expected to work \citep{Armitage11}.
To evaluate the effect of thermal ionization on MRI turbulence, 
correct thermodynamics \citep{Hirose11,Flock13} is crucial since
thermal ionization has strong temperature dependence.
Thermodynamics has another significance in the inner region.
It may regulate the final accretion onto the central star since the local
relation between the surface density $\Sigma$ and the mass accretion rate $\dot{M}$
is uniquely determined for the disc {\it in thermal equilibrium}. 
The relationship $\dot{M} = \dot{M}(\Sigma)$ is called a
thermal equilibrium curve, which is obtained by solving the vertical structure of the disc.
Of particular interest is when the gas temperatures are around the hydrogen ionization temperature $\sim10^4$ K,
where a thermal equilibrium curve generally exhibits bistability.
{\color{\minor}Then, disk annuli are expected to follow a limit-cycle, switching episodically between high and low accretion states.}
Actually, episodic high accretion states of $\dot{M}\sim 10^{-4}M_\odot\text{yr}^{-1}$, called FU Ori outbursts, are
observed in protoplanetary discs, and some authors applied the {\color{\minor}(hydrogen ionization)} disc instability model
(DIM) to explain them \citep{Kawazoe93,Bell94}.

The DIMs 
were originally developed to explain dwarf nova outbursts
\citep[][for a recent review, see \citet{Lasota01}]{Osaki74,Hoshi79,Meyer81,Cannizzo82,Faulkner83,Mineshige83}.
They employ the $\alpha$ prescription of the shear stress {\color{\minor}\citep{Shakura73}}
and the mixing length theory for thermal convection.
Consequently, in the DIMs, a thermal equilibrium curve strongly depends on two free parameters, 
$\alpha$, the ratio of vertically-integrated shear stress to vertically-integrated thermal pressure, and 
the mixing length in terms of the disc scale height, both of which need to be tuned.
On the other hand, \citet{Hirose14} {\color{\minor}(hereafter Paper I)} have derived a thermal equilibrium curve for dwarf nova discs
from {\it the first principles} using radiation MHD simulations with realistic mean opacities and equation of state (EOS).
They have also found that $\alpha$ is enhanced by a factor of several due to convection near the low surface density end of the upper branch of the
thermal equilibrium curve. 
That is consistent with an observational fact that the high accretion state in dwarf novae characteristically
exhibits $\alpha$ of order $0.1$ {\color{\minor}\citep{Smak99,King07}}.

In this paper, using the same methods as {\color{\minor}in Paper I}, we explore
thermodynamics in the inner region of protoplanetary discs
where MRI turbulence is expected to drive accretion.
We also discuss whether FU Ori outbursts can be attributed to the thermal-viscous limit cycle in the framework of the DIMs.
Since we focus on the intrinsic thermodynamics of accretion discs, we don't consider the stellar irradiation here.
Also, we don't include, for simplicity, non-ideal MHD effects and the net vertical magnetic flux. These are planned to be treated in the future works.

This paper is organized as follows. After we briefly describe our methods in Section \ref{sec:methods},
we present the thermal equilibrium curve and the basic thermodynamical properties in Section \ref{sec:results}.
In section \ref{sec:discussion}, we shed light on some points
regarding the thermal stability that were not discussed in {\color{\minor}Paper I}.
The issue of FU Ori outbursts is also discussed in the section. 
Finally, we give summary in Section \ref{sec:summary}.

\section{Methods}\label{sec:methods}

Here, we describe our methods briefly since we employ the same methods
as in {\color{\minor}Paper I} and a complete description is given there.

\subsection{Basic Equations}
The basic equations are ideal MHD equations and frequency-integrated moment equations of radiative transfer:
\begin{gather}
  \frac{\partial\rho}{\partial t} + \nabla\cdot(\rho\bm{v}) = 0, \\
  \frac{\partial(\rho\bm{v})}{\partial t} + \nabla\cdot(\rho\bm{v}\bm{v}) =
  -\nabla p + \frac{1}{4\pi}(\nabla\times\bm{B})\times\bm{B} + \frac{{\kappa}_\text{R}\rho}{c}\bm{F}, \label{eq:motion}\\
  \frac{\partial e}{\partial t} + \nabla\cdot(e\bm{v}) =
  -(\nabla\cdot\bm{v})p - \left(4\pi B(T) - cE\right){\kappa}_\text{P}\rho, \label{eq:energy_gas}\\
  \frac{\partial E}{\partial t} + \nabla\cdot(E\bm{v}) =
  -\nabla\bm{v}:\mathsf{P} + \left(4\pi B(T) - cE\right){\kappa}_\text{P}\rho - \nabla\cdot\bm{F}, \label{eq:energy_rad}\\
  \frac{\partial\bm{B}}{\partial t} - \nabla\times\left(\bm{v}\times\bm{B}\right) = 0, \label{eq:induction}
\end{gather}
where $\rho$ is the gas density, $e$ the gas internal energy, $p$ the gas pressure, $T$ the gas temperature, $E$ the radiation
energy density, $\mathsf{P}$ the radiation pressure tensor, $\bm{F}$ the
radiation energy flux, $\bm{v}$ the velocity field vector, $\bm{B}$ the magnetic
field vector (in CGS emu units), $B(T) = \sigma_\text{B}T^4/\pi$ the Planck
function ($\sigma_\text{B}$, the Stefan--Boltzmann constant), and $c$ the speed
of light. We use a flux-limited diffusion approximation in the radiative
transfer, where $\bm{F}$ and $\mathsf{P}$ are related to $E$ as $\bm{F} =
-(c\lambda(R)/\kappa_\text{R}\rho)\nabla E$ and $\mathsf{P} =
\mathsf{f}(R)E$. Here $\lambda(R) \equiv (2+R)/(6+3R+R^2)$ is a flux limiter
with $R \equiv |\nabla E|/(\kappa_\text{R}\rho E)$, and $\mathsf{f}(R) \equiv
(1/2)(1-f(R))\mathsf{I} + (1/2)(3-f(R))\bm{n}\bm{n}$ is the Eddington tensor
with $f(R) \equiv \lambda(R) + \lambda(R)^2R^2$ and $\bm{n}\equiv\nabla E/|\nabla E|$ \citep{Turner_01}.

The EOS $p = p(e,T)$, Rosseland-mean opacity $\kappa_\text{R}(\rho,T)$, and Planck-mean opacity $\kappa_\text{P}(\rho,T)$
are tabulated beforehand, which are referred in the simulations. 
Since we assume no explicit resistivity (i.e. assume ideal MHD) and no viscosity in the basic equations,
the turbulent dissipation occurs only numerically. The numerically dissipated energies are captured in the form of internal energy in the gas, effectively
resulting in an additional term $q_\text{diss}^+$ in the gas energy
equation (\ref{eq:energy_gas}) \citep{Hirose06}.
In contrast, the radiation damping, which is another dissipation mechanism, is
  fully resolved in the simulations \citep[][See Appendix \ref{sec:raddamping} for details]{Blaes11}. 

\subsection{Numerical Methods}
We use the shearing box approximation to model a local patch of an accretion disc 
as a co-rotating Cartesian frame $(x,y,z)$ with a linearized Keplerian shear flow
$\bm{v} = -(3/2)\Omega x\hat{\bm{y}}$, where the $x$, $y$, and $z$ directions correspond to the
radial, azimuthal, and vertical directions, respectively, and $\hat{\bm{y}}$ is the unit vector in the $y$ direction \citep{Hawley95}.
In this approximation, the right hand side of the equation of motion (\ref{eq:motion}) has extra terms representing the inertial forces,
$-2\Omega\hat{\bm{z}}\times\bm{v} +3\Omega^2x\hat{\bm{x}} - \Omega^2z\hat{\bm{z}}$, 
where $\hat{\bm{x}}$ and $\hat{\bm{z}}$ are the unit vectors in the $x$ and $z$ direction, respectively.
Shearing-periodic, periodic, and outflow boundary conditions are
used for the $x$, $y$, and $z$ boundaries of the box, respectively \citep{Hirose06}.

The basic equations are solved time-explicitly by ZEUS using the
Method of Characteristics--Constrained Transport (MoC--CT)
algorithm except for the radiation--gas energy exchange terms
$\pm(4\pi B - cE){\kappa}_\text{P}\rho$ and the radiative diffusion term
$-\nabla\cdot\bm{F}$. These terms are coupled and solved time-implicitly using
Newton--Raphson iteration \citep{Tomida13} and the multi-grid method with a Gauss--Seidel smoother.

\subsection{Initial Conditions and Parameters}
A shearing box is characterized by a single parameter, the angular velocity $\Omega$, which appears in the inertial force terms and the shearing periodic boundary condition.
We set $\Omega$ to $2.55 \times 10^{-5}$ s$^{-1}$, which is about 250
times smaller than the dwarf nova disc case in {\color{\minor}Paper I.}
The value corresponds to a distance of $R_0 = 5.89\times10^{11}\text{ cm} = 0.0394\text{ AU} = 8.46 R_\odot$ from the central star of $1M_\odot$.
Although $\Omega$ was chosen arbitrarily, the basic features of the thermal equilibrium curve will not
depend on it strongly. (See, for example, Figure 3 in \citet{Kawazoe93}{\color{\minor}, where they show how the thermal equilibrium curve changes with $\Omega$.)}
Therefore, we expect that the radius we chose reasonably represents the inner region of protoplanetary discs.

The initial conditions for gas and radiation are specified by two parameters, the surface density $\Sigma_0$ and 
the effective temperature ${T_\text{eff}}_0$,\footnote{Here, ``$0$'' denotes an initial value.}
assuming the vertical hydrostatic and radiative equilibriums (see
Appendix in {\color{\minor}Paper I}).
The initial magnetic field is a twisted azimuthal flux tube with zero net vertical flux placed
at the centre of the simulation box. The field strength (typically 30 in terms of plasma beta) 
is chosen so that the initial development of MRI is resolved.
The initial velocity field is the linearized Keplerian shear flow, whose $x$ and $z$ components are perturbed by random noise
of 0.5 \% of the local sound velocity. 

The box size is $(1,4,8)$ in terms of the initial disc thickness $h_0$, and the numbers of cells
are (32,64,256), in the $x$, $y$, and $z$ directions, respectively. These values are the same as {\color{\minor}those employed in
{\color{\minor}Paper I}. We will discuss the numerical convergence of our results in Section \ref{sec:assessment}.}  
Different box sizes are also used in some limited runs.
For details of parameters of runs, see Table \ref{table}.

\section{Results}\label{sec:results}

In this section, we present results of our simulations using the same
diagnostics in {\color{\minor}Paper I}.
The diagnostics are based on horizontally-averaged vertical profiles, which were recorded every 0.01
orbits. The horizontally-averaged vertical profile of quantity $f$, for example, is computed as
\begin{align}
f_\text{ave}(z,t) \equiv \dfrac{\int\!\!\int f(x,y,z,t) dxdy}{\int\!\!\int dxdy},\label{eq:ave}
\end{align}
where the integrations are done over the full extent of the box in $x$ and
$y$.

\subsection{Numerical Procedure}\label{sec:procedure}
Our procedure for obtaining a thermal equilibrium curve is as follows --- (1) We start a simulation from the initial condition
determined by a set of the two parameters $(\Sigma_0, {T_\text{eff}}_0)$.
(2) We continue the simulation until the disc patch reaches a thermal equilibrium that lasts at least for $100$ orbits or experiences a thermal runaway.
(3) For the former, we compute time-averaged effective temperature $\bar{T}_\text{eff}$ and surface density $\bar{\Sigma}$. {\color{\minor}Since a small amount of mass can leave a box from the vertical boundaries, $\bar{\Sigma}$ differs from $\Sigma_0$, typically by several percent (See Table \ref{table}).} --- Repeating the above three steps for many sets of $(\Sigma_0, {T_\text{eff}}_0)$, we obtain a thermal equilibrium curve, or the effective temperature as a function of the surface density, $\bar{T}_\text{eff} = \bar{T}_\text{eff}(\bar{\Sigma})$. 

The time-averaged effective temperature and surface density in the above are computed as
\begin{align}
& 2\sigma_\text{B}\bar{T}^4_\text{eff} \equiv \int\left< q^-\right> dz, \\
& \bar{\Sigma} \equiv \int\left<\rho\right> dz.
\end{align}
The cooling rate {\color{\minor}per unit volume} $q^-$ is defined as
\begin{align}
  q^- \equiv \frac{dF_z}{dz} + \frac{d\left((e+E)v_z\right)}{dz},
\end{align}
{\color{\minor}where $F_z$ is the $z$ component of radiation energy flux $\bm{F}$}.
Here and hereafter, the vertical integration $\int dz$ is done for the full extent of $z$.
The brackets $\left<\right>$ denote time-averaging of a horizontally-averaged quantity; for example, 
\begin{align}
\left< f\right>(z) \equiv \dfrac{1}{{\Delta}}{\int_{{\Delta}} f_\text{ave}(z,t) dt},\label{eq:time_averaging}
\end{align}
where averaging is done for a selected period of $\Delta$ in which the disc patch is in a quasi-steady state
and the MRI near the midplane is reasonably resolved. 
The criterion ($= 100$ orbits) for judging a thermal equilibrium of the disc patch was chosen arbitrarily, but, 
as shown in Table \ref{table}, the averaging period ${\Delta}$ is long enough (at least, several) in terms of the thermal time
$\bar{t}_\text{therm}$ computed (a posteriori) as
\begin{align}
  \bar{t}_\text{therm} \equiv \dfrac{\int \left< e+E\right> dz}{\int \left< q^-\right> dz}.
\end{align}

\subsection{Thermal Equilibrium Curve}\label{sec:thermal_equilibrium_curve} 
Figure \ref{fig:s-curve} shows the thermal equilibrium curve of the disc patch, $\bar{T}_\text{eff} = \bar{T}_\text{eff}(\bar{\Sigma})$,
consisting of the thermal equilibrium solutions defined in section \ref{sec:procedure}.
The right axis shows the corresponding mass accretion rate,
\begin{align}
  \bar{\dot{M}} \equiv \dfrac{4\pi}{3}\dfrac{2\sigma_\text{B}\bar{T}_\text{eff}^4}{\Omega^2}.
\end{align}
The color indicates the time-averaged total Rosseland-mean optical thickness $\bar{\tau}_\text{total} \equiv \int\left<\rho\kappa_\text{R}\right> dz$.
Table \ref{table} lists
time-averaged quantities for the runs shown in the figure.

There are two distinct solution branches; the upper branch ($\bar{T}_\text{eff} \gtrsim 5000 \text{ K}$, or $\bar{\dot{M}} \gtrsim 10^{-5}M_\odot\text{ yr}^{-1}$ and $\bar{\Sigma} \gtrsim 6000 \text{ gcm}^{-2} \equiv \Sigma_\text{min}$) and
the lower branch ($\bar{T}_\text{eff} \lesssim 1900 \text{ K}$, or $\bar{\dot{M}} \lesssim 10^{-7}M_\odot\text{ yr}^{-1}$ and $\bar{\Sigma} \lesssim 10000 \text{ gcm}^{-2} \equiv \Sigma_\text{max}$).
For a limited range of surface density ($\Sigma_\text{min} \lesssim \Sigma \lesssim \Sigma_\text{max}$),
there exist, for a single value of $\Sigma$, equilibrium solutions both on the two branches, exhibiting bistability.
The bistability is caused by a strong temperature dependence of Rosseland-mean opacity around
the hydrogen ionization temperature $T \sim 10^4$ K \citep[e.g.][]{Cannizzo93}. That is, given a surface density within the range,
the disc patch can be hot and opaque ($\bar{\tau}_\text{total} \sim 10^{5.5}$) on the upper branch or
cool and less opaque ($\bar{\tau}_\text{total} \sim 10^{2.5}$) on the lower branch.
Here, the Kramers-type opacities are responsible for the former while atomic and molecular opacities are for the latter (see also Figure \ref{fig:denstemp} for
the temperature dependence of opacities).
Note that linearly unstable equilibrium solutions that appear in the DIMs {\color{\minor}\citep[e.g.][]{Kawazoe93}} are excluded in our simulations where nonlinear evolution
of the disc patch is calculated.

The above features of the S-shaped thermal equilibrium curve are qualitatively similar to,
but quantitatively different from the dwarf nova disc case in
{\color{\minor} Paper I}, in which $\Omega$ is about $250$ times larger than here.
First, both the minimum effective temperature on the upper branch ($\sim 5000$K) and
the maximum effective temperature on the lower branch ($\sim 1900$K) are
about 30\% smaller than those in {\color{\minor} Paper I}.
Such trend of the maximum/minimum effective temperatures with $\Omega$ is also observed in the DIMs \citep[e.g. {\color{\minor}Figure 3} in][]{Kawazoe93}.
Second, the bends of the thermal equilibrium curve occur around a surface density 
that is about 60 times larger ($\Sigma \sim 10000$ gcm$^{-2}$).
This is required from the thermal equilibrium condition of a disc patch,
\begin{align}
  2\sigma_\text{B}T_\text{eff}^4 = \frac{3}{2}\Omega\int w_{r\phi}dz = \frac32\Omega\alpha\int p dz \approx \frac32\Omega\alpha T_\text{mid}\Sigma,\label{eq:thermal_equilibrium}
\end{align}
where $w_{r\phi}$ is the shear stress and the $\alpha$ prescription is used in the second equality \citep{Cannizzo93}.
From this relation (${\color{\minor}\sigma_\text{B}}T_\text{eff}^4 \approx \alpha T_\text{mid}\Omega\Sigma$),
$\Sigma$ must increase by a factor of $250\times0.7^4 \sim 60$ to compensate the increase of a factor of $250$ in $\Omega$ and
the decrease of a factor of $0.7$ in $T_\text{eff}$, provided that $\alpha$ and $T_\text{mid}$ don't change much with $\Omega$, which is the case here.
Third, $\Sigma_\text{max}/\Sigma_\text{min}$,
or the relative range of $\Sigma$ for the bistability, is smaller;
it is $\sim 3$ in {\color{\minor} Paper I} while it is $\sim 1.6$ here.

We see another solution branch, the middle branch ($\bar{T}_\text{eff} \sim 3000$ K, $\bar{\Sigma} \sim 10000$ gcm$^{-2}$),
where the total optical thickness ($\bar{\tau}_\text{total} \sim 10^{6}$) is even higher than on the upper branch.
As shown in the following sections, what makes the solutions on this branch different from those {\it having the same $\Sigma$} on the upper branch is convection
being the heat transport mechanism near the midplane.
We observed thermal equilibria for three different initial surface densities,
$\Sigma_0 = 7949$ (ws0858), $8754$ (ws0831 and wa0831), and $10542$ (wb0828 and wc0828)
although the disc patch in ws0858 and wa0831 collapsed eventually (indicated by downward triangles).
{\color{\minor}Two runs, having the same $\Sigma_0$,
slightly differ in the box size and the resolution (see Table \ref{table}).}
We will discuss the thermal stability of these solutions later in section \ref{sec:stability},
together with the two solutions at the high $\Sigma$ end of the lower branch (ws850 and ws854), 
in which the disc patch flared in the end (indicated by upward triangles).

\subsection{Heat Transport} 

Next, we examine heat transport mechanisms in the thermal equilibrium solutions.
In Figure \ref{fig:energyflux}, time and horizontally-averaged vertical profiles of radiative heat flux $\bar{F}^-_\text{rad}(z)$,
advective heat flux $\bar{F}^-_\text{adv}(z)$, and cumulative heating rate $\bar{F}^+_\text{heat}(z)$ are shown
for a typical upper-branch solution (ws0805),
the solution at the low $\Sigma$ end of the upper branch (ws0837), a middle-branch solution (ws0831), and a typical lower-branch solution (ws0800).
Here, the heat fluxes are computed as
\begin{align}
& \bar{F}^-_\text{rad}(z) \equiv \left< F_z\right>,\\
& \bar{F}^-_\text{adv}(z) \equiv \left<(e + E)v_z\right>,\\
& \bar{F}^+_\text{heat}(z)\equiv \int_0^z\left(\left< q_\text{diss}^+\right> + \left<-\left(\nabla\cdot\bm{v}\right)p\right> + \left<-\nabla\bm{v}:\mathsf{P}\right>\right)dz.
\end{align}

From the energy equations (\ref{eq:energy_gas}) and (\ref{eq:energy_rad}), thermal energy balance in a steady state is written as
\begin{align}
q_\text{diss}^+ - \left(\nabla\cdot\bm{v}\right)p -
\nabla\bm{v}:\mathsf{P} = \nabla\cdot\bm{F} +\nabla\cdot\left((e + E)\bm{v}\right),
\label{eq:balance}
\end{align}
where the left hand side is the total heating rate (turbulent dissipation and compressional heating)
and the right hand side is the total cooling rate (radiative diffusion and advection).
{\color{\minor}Note that the numerical dissipation rate $q_\text{diss}^+$ is not explicitly written in equation (\ref{eq:energy_gas}).}
Therefore, it is expected that
\begin{align}
\bar{F}^+_\text{heat}(z) = \bar{F}^-_\text{rad}(z) + \bar{F}^-_\text{adv}(z),
\label{eq:intbalance}
\end{align}
which is the time and horizontally-averaged version of equation (\ref{eq:balance}).
Actually, equation (\ref{eq:intbalance}) holds well in all solutions shown in the figure, confirming thermal equilibrium there.
We note that the compressional heating $- \left(\nabla\cdot\bm{v}\right)p -\nabla\bm{v}:\mathsf{P}$ can act as virtual dissipation when radiative diffusion exists.
{\color{\minor}That is, radiation diffusion leads to photon damping and makes compressive heating irreversible.}
That is really observed in some solutions
though it is a small fraction (10\% or so) of the total dissipation when vertically-integrated (see Appendix \ref{sec:raddamping}).
The same mechanism, the radiation damping {\color{\minor}\citep[Silk damping;][]{Silk68}}, acts
more efficiently in radiation-dominated discs {\color{\minor}\citep{Agol98,Blaes11}}. 

Figure \ref{fig:energyflux} also clearly shows that the main heat transport mechanism differs between the solutions.
In the typical upper-branch and lower-branch solutions (ws0805 and ws0800), radiative diffusion accounts for
heat transport in the entire heights. However, in the solution at the low $\Sigma$ end of the upper branch (ws0837) and the middle-branch solution (ws0831),
it is advection that transports heat near the midplane.
The heat advection is associated with thermal convection that is induced by high Rosseland-mean opacities around $T \sim 10^4$ K.
Actually, squared (hydrodynamic) Brunt-V\"ais\"al\"a frequency $N^2$ computed as
\begin{align}
&\frac{N^2}{\Omega^2} \equiv \frac{1}{\left<\Gamma_1\right>}\frac{d\ln\left< p\right>}{d\ln z} - \frac{d\ln\left<\rho\right>}{d\ln z} - \ln\left<\rho\right>\frac{d\ln\left<\Gamma_1\right>}{d\ln z}
\end{align}
is negative where advection transports heat. (Here, $\Gamma_1$ is the first generalized adiabatic exponent.)
This, in turn, means that
convection does not cancel the super-adiabaticity ({\color{\minor}Paper I}).

In Figure \ref{fig:slice}, snapshots on the $x$-$z$ plane well contrast a ``radiative'' solution (ws0805, $t=71.6$ orbits)
and a ``convective'' solution (ws0837, $t=108.3$ orbits).
{\color{\minor}(These two solutions correspond to the top two panels in Figure \ref{fig:energyflux} respectively.)}
In the convective solution, convective plumes, whose sizes are a fraction of the time-averaged
pressure scale height\footnote{Both the $x$ and $z$ axes in the figure
are normalized with the time-averaged pressure scale height.},
are clearly seen in the panels of specific entropy, 
density and gas temperature. Also, upward (downward) convective plumes well correspond to
upward (downward) advective heat flux ($ev_z$), respectively. Note that the downward and upward convective plumes are highly
asymmetric under the vertical stratification. The convective motion is strong enough to {\color{\minor}greatly} deform the disc patch.
In the radiative solution, on the other hand, 
the disc patch basically keeps its shape, only {\color{\minor}fluctuating because of} the MRI turbulence and the spiral acoustic waves {\color{\minor}\citep{Heinemann09}}.
{\color{\minor}There is no convection,} as indicated by the entropy gradient, and heat transport is done by radiative diffusion
in the direction of the gradient of gas
temperature. {\color{\minor}Note that the energy exchange between gas and
  radiation is so rapid that gas and radiation share a single temperature here.} 

\subsection{Vertical Profiles}
Figure \ref{fig:density} shows time and horizontally-averaged vertical profiles of density and pressures.
Magnetic pressure is always dominant at high altitudes, supporting an extended exponential atmosphere {\color{\minor}\citep[e.g.][]{Hirose11}}, while
it is dominated by thermal pressure near the midplane. These are characteristic features generally seen in stratified
MRI-turbulent discs. Note, however, that plasma beta near the midplane is rather small in the convective solutions;
especially, it is $\sim 10$ at the low $\Sigma$ end of the upper branch (ws0837), which is about ten times
smaller than the characteristic value.
Also, in the convective solutions, density is
peculiarly flat near the midplane. Such flat density profiles are also observed in convective solutions in \citet{Bodo12},
{\color{\minor}where they solved an energy equation with finite thermal diffusivity and a perfect gas EOS.}
Radiation pressure is always smaller than gas and magnetic pressures except in the upper branch solution ws0805,
where it is a fraction of gas pressure and larger than magnetic pressure near the midplane.

Figure \ref{fig:temperature} shows time and horizontally-averaged vertical profiles of gas temperature $T$ and the ionization fraction $f_\text{ion}$.
In the upper-branch solution ws0805, $T > 10^4$ K and $f_\text{ion} = 1$ at almost entire heights.
At the low $\Sigma$ end of the upper branch (ws0837), such complete ionization still holds near the midplane, but
$T$ and $f_\text{ion}$ drop to $3000$ K and $10^{-4}$, respectively, beyond the photospheres.
In the middle branch solution ws0831, $T$ is just below $10^4$ K and $f_\text{ion}$ is around $10^{-1}$ near the midplane, and they
drop to $2000$ K and $10^{-5}$, respectively, beyond the photospheres. In the lower-branch solution ws0800,
$T$ is around $1000$ K and roughly constant due to rather small optical thickness (as indicated by colors in Figure \ref{fig:s-curve}),
and $f_\text{ion}$ is between $10^{-10}$ and $10^{-8}$.

\subsection{Enhancement of $\alpha$}\label{sec:enhanced_alpha}
{\color{\minor}Paper I} found that the time-averaged $\alpha$, computed as
\begin{align}
  \bar{\alpha} = \dfrac{\bar{W}_{xy}}{\bar{P}_\text{therm}} \equiv \dfrac{\int (\left< -B_xB_y\right> + \left<\rho v_x(v_y+3\Omega x/2)\right>)dz}{\int (\left< p\right> + \left< E\right>/3) dz},
\end{align}
is enhanced near the low $\Sigma$ end of the upper branch due to convection.
Such enhancement of $\bar{\alpha}$ is also observed in our simulations, suggesting that it doesn't depend on $\Omega$ of the disc patch.

Figure \ref{fig:correlation} (a) shows $\bar{\alpha}$ as a function of $\bar{\Sigma}$.
For most of the solutions, $\bar{\alpha}$ is $\sim 0.03$, a typical value of the MRI turbulence.
However, $\bar{\alpha}$ increases as $\bar{\Sigma}$ decreases near the low $\Sigma$ end of the upper branch, and exhibits a maximum value of $\sim 0.14$
at the very end (ws0837). 
Also, middle-branch solutions show rather high values of $\bar{\alpha} \sim 0.06$.
The direct reason for the high $\bar{\alpha}$ on the upper branch
can be seen in the trends of $\bar{W}_{xy}$ and $\bar{P}_\text{therm}$ with $\bar{\Sigma}$ (Figure \ref{fig:correlation} b).
The stress $\bar{W}_{xy}$ and pressure $\bar{P}_\text{therm}$ correlate well in the solutions that show $\bar{\alpha}$ of a typical MRI value.
However, near the low $\Sigma$ end of the upper branch, they deviate from the standard trend on the upper branch in opposite ways;
as $\bar{\Sigma}$ decreases $\bar{W}_{xy}$ goes up while $\bar{P}_\text{therm}$ goes down, which makes $\bar{\alpha} = \bar{W}_{xy}/\bar{P}_\text{therm}$ larger.
Physically, the opposite trends of $\bar{W}_{xy}$ and $\bar{P}_\text{therm}$ can be attributed to convection as {\color{\minor}Paper I} argued.
First, $\bar{W}_{xy}$ goes up because vertical convective plumes (Figure \ref{fig:slice})
can create coherent vertical fields that seed the axisymmetric MRI.
Second, $\bar{P}_\text{therm}$ goes down because convection can cool the disc patch more than
the increased stress, or dissipation heats it.
In our case, the second effect is more prominent in enhancing $\alpha$, judging from Figure \ref{fig:correlation} (b).

Next, we examine the condition for convection to enhance $\bar{\alpha}$ in the MRI turbulence.
{\color{\minor}Paper I} introduced, as a barometer of convection, the advective fraction of heat transport computed
as $\bar{f}_\text{adv}\equiv\left<\tilde{f}_\text{adv}\right>$, where
\begin{align}
  \tilde{f}_\text{adv}(t) \equiv \frac{\int \{F^-_\text{adv}\}\sgn(z)\{p_\text{therm}\}dz}{\int\left(\{F^-_\text{adv}\}+\{F^-_\text{rad}\}\right)\sgn(z)\{p_\text{therm}\}dz}.
\end{align}
Here, $p_\text{therm} \equiv p + E/3$ is used as a weight function, and the brackets $\{\}$ denote a boxcar smoothing of a horizontally-averaged quantity for a single orbit.
As shown in Figure \ref{fig:correlation} (c), solutions that show high $\bar{\alpha}$ also show high $\bar{f}_\text{adv}$,
implying strong convection. However, it is not always true the other way around.
For example, the values of $\bar{f}_\text{adv}$ are similar both near the low $\Sigma$ end of the upper branch and on the middle branch,
but $\bar{\alpha}$ is smaller in the {\color{\minor}latter}.
Also, the high-$\Sigma$-end solutions on the lower branch show high $\bar{f}_\text{adv}$ while they show $\bar{\alpha}$ of a typical MRI value.
The discrepancy between $\bar{\alpha}$ and $\bar{f}_\text{adv}$ in the second example
was also discussed in {\color{\minor}Paper I} and they claimed that Mach number of the convective motion needs to be large for $\bar{\alpha}$ to be enhanced.
Here, we check their claim quantitatively by computing the Mach number of the convective motion as follows:
\begin{align}
  \bar{M}_\text{adv} \equiv \left<\dfrac{1}{\int\{p_\text{therm}\}dz}\int \dfrac{\{(e+E)v_z\}\sgn(z)}{\{(e+E)\}\{c_\text{s}\}}\{p_\text{therm}\}dz\right>,
\end{align}
where the sound speed $c_\text{s}$ is defined as $c_\text{s}^2 \equiv \left(\Gamma_1p + 4/3E/3\right)/\rho$.
As shown in Figure \ref{fig:correlation} (d), $\bar{M}_\text{adv}$ actually correlates with $\bar{\alpha}$ much better than $\bar{f}_\text{adv}$.
Therefore, we confirm that convection alone is not responsible for high $\bar{\alpha}$, but its motion needs to be {\it fast} in terms of Mach number{\color{\minor}s reaching $\gtrsim 0.01$}.

{\color{\minor}Paper I} reported that they observed convective/radiative limit cycle near the low $\Sigma$ end of the upper branch, where $\tilde{f}_\text{adv}(t)$
switches between $0$ and $1$ episodically. However, we didn't observe such a clear ``on/off'' limit cycle in our simulations. For example, at the low $\Sigma$
end of the upper branch (ws0837),
convection occurs rather continuously with $\bar{f}_\text{adv} \sim 0.8$ and the standard deviation of $\tilde{f}_\text{adv}(t)$ is $\sim 0.1$.

Although it is true that convection affects the dynamics of turbulence,
MRI seems still primarily responsible for driving the turbulence since the properties of the standard MRI turbulence are retained.
As shown in Figure \ref{fig:stressbz2}, $\bar{W}_{xy}$ well correlates $\int\left< B_z^2\right> dz$,
and $\int\left< -B_xB_y\right> dz \cong 3 \int\left<\rho v_x(v_y+3\Omega x/2)\right> dz$ in all runs \citep{Sano04}.

{\color{\major}
\subsection{Numerical Convergence}\label{sec:assessment}
Dependence on the grid resolution needs to be looked at when specific numerical
values of $\alpha$ are quoted. For example, 
\citet{Fromang07} have shown that the saturation of MRI turbulence in
unstratified shearing box simulations without net vertical magnetic
field and 
explicit dissipation depends on the grid resolution;
$\alpha$ decreases as the resolution is increased. On the other hand,
as for stratified
shearing box simulations (of MRI turbulence without net vertical field
and explicit dissipation)  
like ours, \citet{Hawley11} have proposed
quantitative diagnostics by which the numerical convergence on the
grid resolution can be
assessed. Specifically, the convergence metrics that show clear
resolution dependence and thus are appropriate for the assessment are
the number of cells 
within the fastest-growing MRI wavelength in the $z$ direction, the
same but in the $y$ direction, the ratio of magnetic stress $-B_xB_y/4\pi$ to magnetic energy $\bm{B}^2/8\pi$
and the ratio of $B_x^2$ to $B_y^2$. Here, we compute the four convergence metrics as
\begin{align}
  &Q_z \equiv \dfrac{\int \sqrt{\left<B_z^2\right>/4\pi\left<\rho\right>} \left<p_\text{therm}\right>dz}{\int \left<p_\text{therm}\right> dz}\dfrac{2\pi}{\Omega\Delta z}, \\
  &Q_y \equiv \dfrac{\int \sqrt{\left<B_y^2\right>/4\pi\left<\rho\right>} \left<p_\text{therm}\right>dz}{\int \left<p_\text{therm}\right> dz}\dfrac{2\pi}{\Omega\Delta y}, \\
  &\alpha_\text{mag} \equiv \dfrac{\int \left(\left<-B_xB_y/4\pi\right>/\left<\bm{B}^2/8\pi\right>\right) \left<p_\text{therm}\right>dz}{\int \left<p_\text{therm}\right> dz},\\
  &f_\text{mag} \equiv \dfrac{\int \left(\left<B_x^2\right>/\left<B_y^2\right>\right) \left<p_\text{therm}\right>dz}{\int \left<p_\text{therm}\right> dz},
\end{align}
where the time and horizontally-averaged thermal pressure
$\left<p_\text{therm}\right>$ is used as a weight
function.\footnote{\color{\major}The results are almost unchanged when
  the time and horizontally-averaged density $\left<\rho\right>$ is
  used as a weight function.} 

The results are shown in Figure \ref{fig:assess}. According to
\citet{Hawley11}, numerical convergence is seen in those 
simulations that show $Q_z\gtrsim 10$ and $Q_y\gtrsim 20$,
and $\alpha_\text{mag} \simeq 0.3$--$0.4$ and $f_\text{mag} \gtrsim 0.15$ are signatures of well-developed MRI turbulence.
In our case, $Q_z$ and $Q_y$ meet the criteria of numerical convergence in most of the solutions.
As for the signatures of well-developed MRI turbulence, $\alpha_\text{mag}$ meets the criterion
while $f_\text{mag}$ is mostly $\sim 0.11$, about 25\% smaller
than the criterion. Looking at Figure 5 in \citet{Hawley11}, however,
$f_\text{mag}$ rises as the resolution is increased, but
begins to level off when it exceeds $\sim 0.1$.
Therefore, we may conclude that MRI turbulence is marginally resolved in most of our solutions
while it is perhaps underresolved in the four solutions on the lower branch that show $f_\text{mag} < 0.1$ and $Q_z, Q_y < 10$.
The values of $\alpha (\sim 0.02)$ in the four solutions should be underestimated.

The above convergence metrics may not be directly applied to the solutions in
which convection comes in, and direct resolution
studies as done in \citet{Fromang07} will be needed for them.
However, such resolution studies are highly computationally demanding for our
current radiation MHD simulations, and thus are left for future works.
We note that {\color{\minor}Paper I} has done a limited resolution study for
similar convective solutions to show that the enhancement of $\alpha$
is not sensitive to the grid resolution. 
}

\section{Discussion}\label{sec:discussion}

\subsection{Causes of the S-shape}
In the DIMs, a thermal equilibrium curve is generally S-shaped on the $\Sigma$--$T_\text{eff}$ plane.
That is, two stable solution branches of a positive slope are connected by an unstable solution branch of a negative slope.
One way to see this is to see equation (\ref{eq:thermal_equilibrium}){\color{\minor}, $\sigma_\text{B}T_\text{eff}^4 \approx \alpha T_\text{mid}\Omega\Sigma$, }
which says that the sign of the slope is determined by the dependence of $T_\text{mid}$ on $T_\text{eff}$.
That is, $\Sigma$ usually correlates positively with $T_\text{eff}$, 
but the correlation becomes negative when $T_\text{mid}$ depends on $T_\text{eff}$ more strongly than $T_\text{eff}^4$ \citep{Cannizzo93}. 
The dependence of $T_\text{mid}$ on $T_\text{eff}$ can be actually examined in our simulations.
As shown in Figure \ref{fig:tmidteff}, it is always weaker than $T_\text{eff}^4$ on the solution branches, including the middle branch, while
it becomes almost critical near the low $\Sigma$ end of the upper branch. 
Therefore, our results also imply that the disc patch is thermally unstable when
the dependence of $T_\text{mid}$ on $T_\text{eff}$ is stronger than $T_\text{eff}^4$.


The dependence of $T_\text{mid}$ on $T_\text{eff}$ can also be qualitatively
discussed with time and horizontally-averaged 
vertical profiles of the solutions $(\left< T\right>(z),\left<\rho\right>(z))$ plotted on the $T$--$\rho$ plane \citep[][]{Pojmanski86}.
As shown in Figure \ref{fig:denstemp},
two gaps divide a bunch of profile curves into three groups, which, from the left to the right,
correspond to the lower branch, the middle branch, and the upper branch.
The slope of a profile curve is always positive for all solutions, indicating that $\left<{T}\right>$ and $\left<{\rho}\right>$ are monotonically increasing
from the surface toward the midplane. We don't see
  any density inversion that is seen in some convective solutions in the DIMs \citep{Pojmanski86}.
The profile curves of the lower branch solutions are nearly vertical since the total optical thickness $\bar{\tau}_\text{total}$ is relatively small. 
On the other hand, the profile curves of the solutions near the low $\Sigma$ end of the upper branch are almost horizontal ($\rho \sim$ const.) near the midplane
due to convection (as we saw in Figure \ref{fig:density}) while they are almost vertical ($T \sim$ const.) at high altitudes because of low opacities.
Consequently, the gap between the upper-branch and the lower-branch profile curves
is larger in $T_\text{mid}$ than in $T_\text{eff}$.
In other words, a small change in $T_\text{eff}$ results in a large change in $T_\text{mid}$, or
$T_\text{mid}$ strongly depends on $T_\text{eff}$ between the upper and lower branches.
Note that the profile curves of the middle branch solutions are nearly parallel,
indicating that $T_\text{mid}$ doesn't strongly depend on $T_\text{eff}$ (as seen in Figure \ref{fig:tmidteff}).

Figure \ref{fig:denstemp} also tells us about what is happening at the low (high) $\Sigma$ end of the upper (lower) branch, respectively.
Near the low $\Sigma$ end of the upper branch, as $\Sigma$ decreases, a convective region appears away from the midplane.
The convective region finally extends to the midplane at the end of the branch, where the disc patch loses a thermal
equilibrium \citep{Pojmanski86,Cannizzo93}. Note that the convective region on the upper branch well correlates with
the region of high Rosseland-mean opacity (panel (b)), 
implying that the convection is induced by the high opacities.
On the other hand, near the high $\Sigma$ end of the lower branch, convection appears at the midplane as $\Sigma$ increases.
Since the opacity is still low there (panel (b)),
the main cause of convection may be low $\Gamma_1$ (panel (c)),
which reduces the adiabatic temperature gradient. 
The disc patch seems to lose an equilibrium when the opacity source switches from atoms/molecules to H$^-$ at the end of the branch.

\subsection{Thermal Stability of Equilibrium Solutions}\label{sec:stability}
When the cooling rate $Q^-$ depends on temperature $T$ more strongly than the heating rate $Q^+$ does, 
\begin{align}
  \frac{d\ln Q^+}{d\ln T} < \frac{d\ln Q^-}{d\ln T},\label{eq:stability_condition}
\end{align}
the disc patch is thermally (linearly) stable. The condition can be written as
\begin{align}
  \frac{d\log Q^-}{d\log Q^+} > 1.\label{eq:stability_condition2}
\end{align}
We examine this condition for selected thermal equilibrium solutions including those exhibit a thermal runaway in the end.
Figure \ref{fig:stability} shows time trajectories of the selected solutions on the $\log \tilde{Q}^+$--$\log \tilde{Q}^-$ plane.
Here, $\tilde{Q}^+$ and $\tilde{Q}^-$ are {\color{\minor}short-term average,} vertically-integrated heating and cooling rates,
\begin{align}
  &\tilde{Q}^+(t) \equiv \int\left\{q_\text{diss}^+ - \left(\nabla\cdot\bm{v}\right)p - \nabla\bm{v}:\mathsf{P}\right\} dz,\\
  &\tilde{Q}^-(t) \equiv \int\left\{\dfrac{dF}{dz} + \dfrac{(e + E)v_z}{dz}\right\} dz.
\end{align}
The centres of these trajectories are roughly on
the line $\log \tilde{Q}^- = \log \tilde{Q}^+$, showing that the solutions are thermally-stable in a statistical sense. 

First, we see typical solutions on the upper branch (ws0803{\color{\minor}, the top panel})
and on the lower branch (ws0800{\color{\minor}, the bottom panel}), both of which are
radiative solutions.
If we look at the trajectory curve closely, it is almost horizontal {\color{\minor}on short timescales} due to a rapid stochastic variation of $\tilde{Q}^+(t)$.
Even in a statistical sense, the trajectory is roughly aligned with ${d\log \tilde{Q}^-}/{d\log \tilde{Q}^+} = 1$, rather
than satisfying the condition (\ref{eq:stability_condition2}). Therefore, the simple linear theory does not apply here.
Next, we see the trajectories of the convective solutions on the upper branch (ws0837) and the middle branch (wa0831 and ws0831).
They are more compact than those of the radiative solutions, and thus the slope is less clear.
Also, it is not clear from their similar trajectories why the disc patch finally collapsed only in wa0831, and not in the other two cases.
On the other hand, the trajectory of the solution at the high $\Sigma$ end of the lower branch (ws0850), which also experienced a runaway finally,
looks different from those mentioned in the above, and its slope is apparently less than unity,
indicating thermal {\color{\minor}instability}.

The slope of the trajectory may be quantitatively evaluated as $\delta\log\tilde{Q}^-/\delta\log\tilde{Q}^+$,
the ratio of the standard deviation of $\log \tilde{Q}^-(t)$ to that of $\log \tilde{Q}^+(t)$.
In Figure \ref{fig:stability2}, we plot the ratio as a function of $\bar{\Sigma}$ to see the trend among the equilibrium solutions. 
The ratio is almost unity for most of the solutions (as we see for ws0803 and ws0800 in the above).
However, it is obviously smaller 
for solutions near the ends of the upper and lower branches and on the middle branch.
It is especially small (less than $0.7$)
for the solutions that finally experienced a thermal runaway (as indicated by triangles), which suggests that
the ratio can be an indicator of {\color{\minor}``fragility'' of the thermal equilibrium \citep{Hirose09,Jiang13}}. 
Other solutions {\color{\minor}with $\tilde{Q}^-(t)$ varying less than the $\tilde{Q}^+(t)$} stayed in equilibrium {\color{\minor}up to} the end of the simulations, but
{\color{\minor}nevertheless} are potentially unstable since {\color{\minor}$\tilde{Q}^+(t)$ varies
stochastically and $\tilde{Q}^-(t)$ does not always respond on the same timescale.
So, if $\tilde{Q}^+(t)$ wanders around the equilibrium or returns quickly after
departing, then $\tilde{Q}^-(t)$ can respond and the system is stable.  However, a
sufficiently large and long-lasting excursion in $\tilde{Q}^+(t)$ builds up a heat
excess or deficit so big that the patch of disk does not return to the
equilibrium.  Thanks to this fragility}, the ends of the upper and
lower branches are not well-defined \citep{Latter12}.
{\color{\minor}Another consequence is that the middle branch is long-lasting in some
runs, but not in others with similar parameters, as also observed in
Paper I.}

The stability of the middle branch solutions can be discussed from another viewpoint. 
In the DIMs, it is explained that the portion of a negative slope of the S-shaped thermal equilibrium curve is thermally unstable
due to a strong temperature dependence of H$^-$ opacity \citep[e.g.][]{Kato08}.
That is, using $\kappa_\text{R} \sim \rho^{1/2}T^9$ for H$^-$, the cooling rate anti-correlates with the midplane temperature as
\begin{align}
  Q^- \sim T_\text{eff}^4 \sim \frac{T_\text{mid}^4}{\kappa_\text{R}\Sigma} \sim \frac{T_\text{mid}^{-19/4}}{\Sigma^{3/2}},\label{eq:cool_temperature_dependence}
\end{align}
while $Q^+ \sim \Sigma T_\text{mid}$ assuming the $\alpha$ prescription, violating the thermally stable condition (\ref{eq:stability_condition}).
This is, however, true when the disc patch is cooled only
by radiative diffusion in which $\sigma_\text{B}T_\text{eff}^4 \approx 4acT_\text{mid}^4/3\kappa_\text{R}\Sigma$ ($a$ is the radiation constant).
To put it {\color{\minor} the other way round}, another cooling mechanism, or convection, may destroy the condition (\ref{eq:cool_temperature_dependence}), stabilizing the disc patch.
Actually, all the middle branch solutions have $\Gamma_1$ of the smallest value ($\sim 1.1$) near the midplane, implying
that convection has been induced most strongly there.

\subsection{FU Ori Outbursts}
The FU Ori outbursts have characteristic amplitudes of $\sim 6$ magnitudes, and the durations of
the outburst and the quiescence are evaluated as $\sim 10^2$ years and $\sim 10^3$ years, respectively \citep{Hartman96}.
Some authors have applied the DIMs to attribute the outburst cycle to the thermal-viscous limit cycle \citep{Kawazoe93,Bell94}.
In the DIMs, $\alpha$ is a free parameter and is chosen so that the observed outburst cycle is reproduced.
Choosing the same $\alpha$ both on the upper and lower branches generally produces a smaller amplitude and a smaller duty cycle
(= the ratio of the outburst duration to the quiescence duration) than the observations.
Therefore, the upper branch of a high $\alpha$ and the lower branch of a low $\alpha$
are usually combined. {\color{\minor}(This is because a thermal equilibrium curve generally shifts to lower right as $\alpha$ decreases.)}
Specifically, $\alpha_\text{hot}$ needs to be larger than $\alpha_\text{cool}$
by a factor of {\color{\minor}10} to reproduce the observed amplitude and duty cycle.
{\color{\minor}Here, $\alpha_\text{hot}$ and $\alpha_\text{cool}$ denotes $\alpha$ on
  the upper and lower branches, respectively, which is a common notation in the DIMs.}
Also, to explain the absolute value of the outburst duration, it is required that 
$\alpha_\text{hot}\sim 10^{-3}$.

\begin{figure}
  \centering
    \includegraphics[width=15cm]{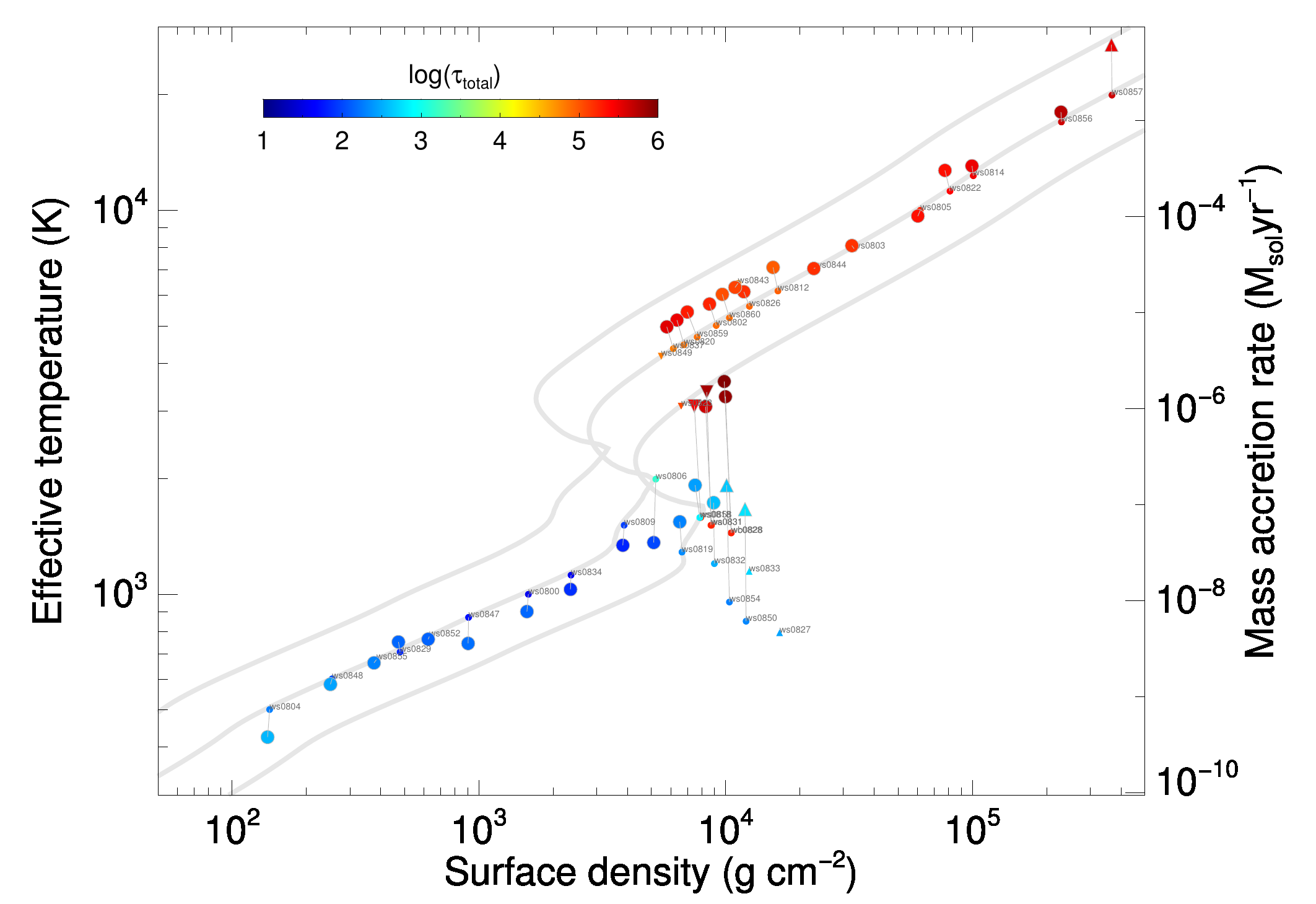}
    \caption{
      Time-averaged effective temperature $\bar{T}_\text{eff}$ vs. surface density $\bar{\Sigma}$ of runs that reached a thermal equilibrium (large circles).
      Large upward (downward) triangles indicate solutions that showed runaway heating (cooling) in the end, respectively.
      The initial condition of each run $({T_\text{eff}}_0,\Sigma_0)$ is shown as a small circle, which is connected to $(\bar{T}_\text{eff},\bar{\Sigma})$ by
      a gray line. Small upward (downward) triangles indicate the initial conditions of runs
      that didn't reach a thermal equilibrium and showed runaway heating (cooling), respectively. 
      Colors represent the time-averaged Rosseland-mean optical thickness $\bar{\tau}_\text{total}$.
      Gray thick curves are thermal equilibria produced by a DIM {\it without convection} {\color{\major}(see Appendix in Paper I)},
      for $\alpha = 0.1$, $0.03$, and $0.01$, from the top to the bottom.
      The name of each run is printed {\color{\minor}next to its $({T_\text{eff}}_0,\Sigma_0)$ point} with a small (but scalable) font for reference.}
  \label{fig:s-curve}
\end{figure}

By contrast, we have uniquely identified from the first principles simulations that $\alpha_\text{hot} \sim {\color{\minor}0.14}$ (near the low $\Sigma$ end)
and $\alpha_\text{cool} \sim 0.03$ as well as $\Sigma_\text{max}/\Sigma_\text{min} \sim 1.6$.
{\color{\minor}Comparing with, for example, Figure 3 in
  \citet{Bell94}, $\alpha_\text{hot} / \alpha_\text{cool} \sim
        {\color{\minor}5}$ is smaller by a factor of two 
  while $\Sigma_\text{max}/\Sigma_\text{min}\sim 1.6$ is an order of
  magnitude smaller. One of the consequences of the smaller surface
  density contrast across the bistability would be a smaller amplitude of
  the outbursts. Actually, if }
 we supposed a thermal-viscous limit cycle on our thermal equilibrium curve (neglecting the subtle middle branch) in Figure \ref{fig:s-curve},
the mass accretion rate would be ${\color{\minor}\sim} 10^{-5}M_\odot\text{yr}^{-1}$ in the outburst and
${\color{\minor}\sim} 10^{-7}M_\odot\text{yr}^{-1}$ in the quiescence, the former being an order of magnitude smaller than that suggested by observations.
Furthermore, the duration of the outburst would be too short with our $\alpha_\text{hot}\sim {\color{\minor}0.14}$;
it is two orders of magnitude larger than that chosen in the DIMs to reproduce the outburst lasting $\sim 10^2$ years.
Thus, we may conclude that the thermal-viscous limit cycle alone would not explain
the FU Ori outbursts. Note, however, that the above discussion is true provided 
that the turbulence is driven by MRI both on the upper and lower branches.
As we saw in Figure \ref{fig:temperature}, temperatures in the lower branch solutions are typically $\sim 10^3$ K.
Therefore, the lower branch solutions may need to be reconsidered taking account of the non-ideal MHD effects
as well as the stellar irradiation that could reduce the effects.

Recently, another model for FU Ori Outbursts has been proposed \citep{Armitage01,Zhu09,Martin11}:
Since temperatures in protoplanetary discs (except the inner region) would be too low for good coupling between gas and magnetic field,
a dead zone is expected where MRI turbulence is suppressed and the accreting gas piles up.
Then, eventually, the gravitational instability would drive turbulence, which dissipates to heat the disc while the disc is
cooled by radiation \citep{Gammie01,Shi14}.
Such thermal equilibria may compose another solution branch, the ``gravito-turbulence'' branch,
below the classical S-shaped thermal equilibrium curve on the $\Sigma$--$T_\text{eff}$ plane.
Then, a limit cycle switching the gravito-turbulence branch and the MRI-turbulence branch
would become possible \citep{Martin11}.
This gravo-magneto limit cycle is shown to be reasonably consistent with the observations of FU Ori outbursts \citep{Zhu09}.

We comment on the rightmost two solutions on the upper branch (ws0856 and ws0857).
In these solutions, the shearing box approximation may not be valid since the pressure scale height $\bar{h}_\text{p}$ of the disc patch is
comparable or larger than the distance from the star $R_0$ (See Table \ref{table}). 
Nevertheless, it might be worth to note that the ratio of vertically-integrated radiation pressure to vertically-integrated gas pressure exceeds unity
($\sim 1.7$ in ws0856 and $\sim 2.9$ in ws0857).
They imply that the inner region of protoplanetary discs can be radiation-pressure dominated when
$\dot{M} \gtrsim 10^{-3} M_\odot\text{yr}^{-1}$, which is an order of magnitude higher than that of the FU Ori outbursts though.
Also, we note that a long-term build-up of heat is observed
in ws0857 (while it is not in ws0856),
which in fact is indicated by a relatively small value of $\delta\log\tilde{Q}^-/\delta\log\tilde{Q}^+$ in Figure \ref{fig:stability2} (see discussion \ref{sec:stability}).

{\color{\major}
  Finally, we note that the $\alpha$ values obtained in shearing box
  simulations are generally those of (local) {\it quasi-steady}
  states. On the other hand, $\alpha$ values cannot
  be deduced from observations of steady accretion discs, and they are 
  usually evaluated utilizing {\it transient} phenomena of the viscous
  timescale $t_\text{vis}$ at radius $R$, using the following relation
  \citep[e.g.][]{Kotko12}, 
  \begin{align}
    t_\text{vis} \sim \frac{R^2}{\nu_\text{vis}} \sim
    \frac{R^2\Omega}{\alpha c_\text{s}^2},
  \end{align}
  where $\nu_\text{vis} (\sim\alpha c_\text{s}^2/\Omega)$ is the viscosity coefficient and $c_\text{s}$
  is the sound speed.\footnote{\color{\major}Since $c_\text{s}$ can be
    expressed as a function of $R$ based on the {\it steady} $\alpha$ model
    \citep{Shakura73}, $\alpha$ can be evaluated once the viscous
    timescale $t_\text{vis}$ and the radius $R$ are given.} One needs
  to keep the above difference in mind when one compares $\alpha$ values between shearing
  box simulations and observations
  \citep{King07,Cannizzo12}. Actually, \citet{Sorathia12} showed that
  the $\alpha$ values in the initial transient can be orders of
  magnitude larger than those in the quasi-steady state 
  in their global (unstratified, isothermal) simulations.
}

\section{Summary}\label{sec:summary}

Using 3D radiation MHD simulations with realistic mean opacities and EOS, we explored thermodynamics
in the inner part of protoplanetary discs where MRI turbulence is expected.
Basic thermal properties are similar to those of dwarf nova discs in {\color{\minor}Paper I}
in spite of a two-orders-of-magnitude smaller $\Omega$.
The thermal equilibrium curve consists of the upper, lower, and middle branches.
The upper (lower) branch corresponds to a hot (cool) and optically {\color{\minor}very (moderately) thick} disc patch, respectively,
while the middle branch is characterized by convective energy transport near the midplane.
Convection is also the major energy transport mechanism near the low $\Sigma$ end of the upper branch.
There, convective motion is fast in terms of Mach number{\color{\minor}s reaching $\gtrsim 0.01$}, which enhances both MRI turbulence and cooling,
raising $\alpha$ up to $0.14$ that other wise is $\sim 0.03$.
The enhancement of $\alpha$ due to convection near the low $\Sigma$ end of the upper branch
was also found in {\color{\minor}Paper I}, and thus seems robust regardless of $\Omega$ of the disc patch.

We examined causes of the S-shape of the thermal equilibrium curve
based on the dependence of $T_\text{mid}$ on $T_\text{eff}$ 
and the time and horizontally-averaged vertical profiles plotted on the $T$-$\rho$ plane.
Then, we discussed thermal stability of the equilibrium solutions
based on their trajectories on the $\log Q^+$--$\log Q^-$ plane.
We also compared our results with the DIMs {\color{\minor}used to explain} FU Ori outbursts \citep{Kawazoe93,Bell94}.
  {\color{\minor}Although the thermal equilibrium curve in our results also exhibits bistability,
    the surface density contrast across the bistability is an order of magnitude smaller, 
    and the stress-to-pressure ratios in both upper and lower branches are two orders of magnitude greater,
    than those favored in the DIMs.}
  {\color{\minor}It therefore appears likely that} FU Ori outbursts {\color{\minor}are not due solely to a} thermal-viscous limit cycle
  {\color{\minor}resulting from} accretion driven by {\color{\minor}local} MRI turbulence.
Instead, the limit cycle switching the gravito-turbulence branch and the MRI-turbulence branch
may be working there \citep{Armitage01,Zhu09,Martin11}.

We note some caveats in our results. We ignored the stellar irradiation, although it is major heating source in protoplanetary discs,
to concentrate on the intrinsic thermodynamics of accretion discs.
Also, non-ideal MHD effects, which we didn't include, might be important in the lower (cool) branch solutions.
We are planning to take account of those as well as the net vertical magnetic flux in the future works.

\section*{Acknowledgments}
We thank {\color{\minor}N. J. Turner, }O. Blaes and J. H. Krolik for useful comments and suggestions.
{\color{\minor}We also thank the anonymous referee for his/her valuable comments for improving the manuscript.}
This work was supported by JSPS KAKENHI Grant 24540244 and 26400224,
joint research project of ILE, Osaka University.
Numerical simulations were carried out on Cray XC30 at CfCA, National Astronomical Observatory of Japan and
SR16000 at YITP in Kyoto University.


\clearpage

\begin{table}
  \centering
  \caption{List of runs.}
  \scalebox{0.7}{
    \begin{tabular}{llrrrrrrrrrrrrrrrrrr}
      \hline
      \multicolumn{1}{c}{run} &
      &
      \multicolumn{1}{c}{$\Sigma_0$} &
      \multicolumn{1}{c}{${T_\text{eff}}_0$} &
      \multicolumn{1}{c}{$\alpha_0$} &
      \multicolumn{1}{c}{$h_0$} &
      \multicolumn{1}{c}{$\bar{\Sigma}$} &
      \multicolumn{1}{c}{$\bar{T}_\text{eff}$} &
      \multicolumn{1}{c}{$\bar{\alpha}$} &
      \multicolumn{1}{c}{$\bar{\tau}_\text{total}$} &
      \multicolumn{1}{c}{$N_x$} &
      \multicolumn{1}{c}{$N_y$} &
      \multicolumn{1}{c}{$N_z$} &
      \multicolumn{1}{c}{$L_x'$} &
      \multicolumn{1}{c}{$L_y'$} &
      \multicolumn{1}{c}{$L_z'$} &
      \multicolumn{1}{c}{$L_x/\bar{h}_\text{p}$} &
      \multicolumn{1}{c}{$\bar{h}_\text{p}/R_0$} &
      \multicolumn{1}{c}{$\bar{t}_\text{therm}$} &
      \multicolumn{1}{c}{${\Delta}$}\\
      \hline
      ws0857 & Ue & 367337 & 19952 & 0.0292 & 7.29e+11 & 365870 & 26982 & 0.0250 &  334485 &  32 &  64 & 384 &  1.0 &  4.0 & 12.0 &  1.03 &  1.20 & 12.13 &  8.25 \\
ws0856 &  U & 230461 & 16982 & 0.0294 & 6.49e+11 & 228452 & 18022 & 0.0162 &  547172 &  32 &  64 & 384 &  0.7 &  2.8 &  8.4 &  1.07 &  0.72 & 16.82 &  5.95 \\
ws0814 &  U & 101199 & 12302 & 0.0295 & 6.39e+11 & 119438 & 13658 & 0.0264 &  312873 &  32 &  64 & 256 &  0.6 &  2.4 &  4.8 &  1.20 &  0.54 & 10.29 &  8.74 \\
ws0822 &  U &  81609 & 11220 & 0.0295 & 4.80e+11 &  77389 & 12703 & 0.0363 &  222433 &  32 &  64 & 256 &  0.5 &  2.0 &  4.0 &  0.95 &  0.43 &  7.79 & 11.55 \\
ws0805 &  U &  61523 & 10000 & 0.0293 & 4.44e+11 &  60080 &  9663 & 0.0256 &  217544 &  32 &  64 & 256 &  0.5 &  2.0 &  4.0 &  1.12 &  0.34 & 11.60 &  8.62 \\
ws0803 &  U &  33289 &  7943 & 0.0294 & 3.73e+11 &  32472 &  8090 & 0.0321 &  145031 &  32 &  64 & 256 &  0.5 &  2.0 &  4.0 &  1.09 &  0.29 & 10.14 &  9.86 \\
ws0844 &  U &  23176 &  7079 & 0.0312 & 3.57e+11 &  22759 &  7054 & 0.0306 &  155910 &  32 &  64 & 256 &  0.5 &  2.0 &  4.0 &  1.16 &  0.26 & 10.94 &  8.22 \\
ws0827 &  e &  16327 &   794 & 0.0010 & 3.71e+10 &  --- &  --- &  --- &  --- &  32 &  64 & 256 &  0.5 &  2.0 &  4.0 & --- & --- & --- & --- \\
ws0812 &  U &  16273 &  6165 & 0.0293 & 2.87e+11 &  15563 &  7102 & 0.0490 &   95897 &  32 &  64 & 256 &  0.5 &  2.0 &  4.0 &  0.95 &  0.26 &  7.16 & 16.76 \\
ws0826 &  U &  12530 &  5623 & 0.0296 & 2.56e+11 &  11847 &  6136 & 0.0470 &  156332 &  32 &  64 & 256 &  0.5 &  2.0 &  4.0 &  1.00 &  0.22 &  8.12 & 12.31 \\
ws0833 &  e &  12389 &  1148 & 0.0029 & 4.17e+10 &  --- &  --- &  --- &  --- &  32 &  64 & 256 &  0.5 &  2.0 &  4.0 & --- & --- & --- & --- \\
ws0850 & Le &  12133 &   851 & 0.0017 & 3.68e+10 &  11974 &  1665 & 0.0185 &     574 &  32 &  64 & 256 &  0.5 &  2.0 &  4.0 &  1.12 &  0.03 & 28.45 &  2.81 \\
ws0843 &  U &  11193 &  6456 & 0.0551 & 2.96e+11 &  10912 &  6299 & 0.0559 &  123578 &  32 &  64 & 256 &  0.5 &  2.0 &  4.0 &  1.14 &  0.22 &  6.78 & 16.23 \\
wb0828 &  M &  10542 &  1445 & 0.0029 & 4.55e+10 &   9977 &  3270 & 0.0548 &  766821 &  32 &  64 & 256 &  0.9 &  3.6 &  7.2 &  1.08 &  0.06 & 10.73 &  9.32 \\
wc0828 &  M &  10542 &  1445 & 0.0029 & 4.55e+10 &   9868 &  3583 & 0.0665 & 1387583 &  32 &  64 & 256 &  1.0 &  4.0 &  8.0 &  1.09 &  0.07 &  9.70 & 15.46 \\
ws0860 &  U &  10342 &  5248 & 0.0296 & 2.35e+11 &   9688 &  6041 & 0.0561 &  108191 &  32 &  64 & 256 &  0.5 &  2.0 &  4.0 &  0.92 &  0.22 &  7.09 & 14.10 \\
ws0854 & Le &  10267 &   954 & 0.0030 & 3.74e+10 &  10086 &  1922 & 0.0313 &     426 &  32 &  64 & 256 &  0.5 &  2.0 &  4.0 &  1.01 &  0.03 & 22.06 &  4.53 \\
ws0802 &  U &   9123 &  5011 & 0.0296 & 2.21e+11 &   8598 &  5700 & 0.0656 &  177212 &  32 &  64 & 256 &  0.5 &  2.0 &  4.0 &  0.99 &  0.19 &  6.56 & 14.47 \\
ws0832 &  L &   9036 &  1202 & 0.0055 & 3.98e+10 &   8933 &  1731 & 0.0251 &     386 &  32 &  64 & 256 &  0.5 &  2.0 &  4.0 &  1.20 &  0.03 & 21.69 &  4.61 \\
ws0831 &  M &   8754 &  1513 & 0.0030 & 4.86e+10 &   8294 &  3081 & 0.0584 &  429956 &  32 &  64 & 256 &  0.7 &  2.8 &  5.6 &  0.97 &  0.06 & 10.18 &  9.82 \\
wa0831 & Mc &   8754 &  1513 & 0.0030 & 4.86e+10 &   8378 &  3364 & 0.0689 &  659101 &  32 &  64 & 256 &  1.0 &  4.0 &  8.0 &  1.29 &  0.06 &  8.68 & 11.52 \\
ws0858 & Mc &   7949 &  1584 & 0.0030 & 5.30e+10 &   7467 &  3093 & 0.0675 &  291339 &  32 &  64 & 256 &  0.7 &  2.8 &  5.6 &  1.06 &  0.06 &  8.62 & 11.61 \\
ws0818 &  L &   7852 &  1584 & 0.0099 & 4.07e+10 &   7517 &  1924 & 0.0347 &     276 &  32 &  64 & 256 &  0.5 &  2.0 &  4.0 &  1.11 &  0.03 & 19.44 &  5.14 \\
ws0859 &  U &   7609 &  4677 & 0.0294 & 2.02e+11 &   6987 &  5434 & 0.0842 &  190815 &  32 &  64 & 256 &  0.5 &  2.0 &  4.0 &  1.04 &  0.16 &  5.51 & 18.15 \\
ws0820 &  U &   6776 &  4466 & 0.0294 & 1.90e+11 &   6346 &  5178 & 0.1081 &  324784 &  32 &  64 & 256 &  0.5 &  2.0 &  4.0 &  1.16 &  0.14 &  4.87 & 20.55 \\
ws0819 &  L &   6671 &  1288 & 0.0099 & 3.68e+10 &   6514 &  1544 & 0.0301 &     197 &  32 &  64 & 256 &  0.5 &  2.0 &  4.0 &  1.23 &  0.03 & 13.96 &  7.16 \\
ws0823 &  c &   6535 &  3090 & 0.0099 & 1.37e+11 &  --- &  --- &  --- &  --- &  32 &  64 & 256 &  0.5 &  2.0 &  4.0 & --- & --- & --- & --- \\
ws0837 &  U &   6153 &  4365 & 0.0312 & 1.88e+11 &   5771 &  4971 & 0.1314 &  363722 &  32 &  64 & 256 &  0.5 &  2.0 &  4.0 &  1.33 &  0.12 &  4.45 & 25.84 \\
ws0849 &  c &   5517 &  4168 & 0.0312 & 1.75e+11 &  --- &  --- &  --- &  --- &  32 &  64 & 256 &  0.5 &  2.0 &  4.0 & --- & --- & --- & --- \\
ws0806 &  L &   5210 &  1995 & 0.0296 & 3.94e+10 &   5109 &  1364 & 0.0253 &     100 &  32 &  64 & 256 &  0.5 &  2.0 &  4.0 &  1.44 &  0.02 & 12.49 &  8.00 \\
ws0809 &  L &   3876 &  1513 & 0.0294 & 3.24e+10 &   3829 &  1342 & 0.0330 &      66 &  32 &  64 & 256 &  0.5 &  2.0 &  4.0 &  1.22 &  0.02 &  8.92 & 11.21 \\
ws0834 &  L &   2346 &  1122 & 0.0288 & 3.39e+10 &   2349 &  1030 & 0.0229 &      78 &  32 &  64 & 256 &  0.5 &  2.0 &  4.0 &  1.39 &  0.02 & 11.53 &  8.68 \\
ws0800 &  L &   1568 &  1000 & 0.0287 & 3.20e+10 &   1564 &   901 & 0.0221 &     145 &  32 &  64 & 256 &  0.5 &  2.0 &  4.0 &  1.37 &  0.02 & 11.90 &  8.40 \\
ws0847 &  L &    908 &   870 & 0.0309 & 2.94e+10 &    903 &   744 & 0.0203 &     159 &  32 &  64 & 256 &  0.5 &  2.0 &  4.0 &  1.32 &  0.02 & 12.78 &  6.26 \\
ws0852 &  L &    625 &   776 & 0.0307 & 2.73e+10 &    622 &   764 & 0.0328 &     141 &  32 &  64 & 256 &  0.5 &  2.0 &  4.0 &  1.23 &  0.02 &  7.91 & 12.64 \\
ws0829 &  L &    475 &   707 & 0.0290 & 2.57e+10 &    471 &   751 & 0.0412 &     145 &  32 &  64 & 256 &  0.5 &  2.0 &  4.0 &  1.16 &  0.02 &  6.36 & 18.88 \\
ws0855 &  L &    384 &   676 & 0.0310 & 2.51e+10 &    375 &   662 & 0.0319 &     199 &  32 &  64 & 256 &  0.5 &  2.0 &  4.0 &  1.16 &  0.02 &  8.17 & 12.24 \\
ws0848 &  L &    252 &   602 & 0.0309 & 2.34e+10 &    250 &   583 & 0.0301 &     285 &  32 &  64 & 256 &  0.5 &  2.0 &  4.0 &  1.11 &  0.02 &  8.63 & 11.59 \\
ws0804 &  L &    142 &   501 & 0.0296 & 2.02e+10 &    139 &   424 & 0.0227 &     351 &  32 &  64 & 256 &  0.5 &  2.0 &  4.0 &  1.17 &  0.01 & 11.26 & 13.32 \\

      \hline
    \end{tabular}
  }
  The characters in the second column represents the outcome: U = upper branch, M = middle branch, L = lower branch, c = runaway cooling, and e = runaway heating.
  The units of surface
  densities ($\Sigma_0$ and $\bar{\Sigma}$), effective temperatures
  (${T_\text{eff}}_0$ and $\bar{T}_\text{eff}$), height ($h_0$), and thermal time ($t_\text{th}$) are, respectively, gcm$^{-2}$, K, cm, and orbit. $L_x'$, $L_y'$,
  and $L_z'$ are the box size (normalized by $h_0$), and $N_x$, $N_y$, and $N_z$ are the number of cells,
  in the $x$, $y$, and in $z$ directions, respectively. 
  ${\Delta}$ is the period of time-averaging used in the diagnostics, normalized by $\bar{t}_\text{therm}$. 
  The pressure scale height is computed as $\bar{h}_\text{p} \equiv \int\left< p_\text{therm}\right> dz / 2\max(\left< p_\text{therm}\right>) $.
  \label{table}
\end{table}
\clearpage

\clearpage

\begin{figure}
  \centering
  \includegraphics[width=6.5cm]{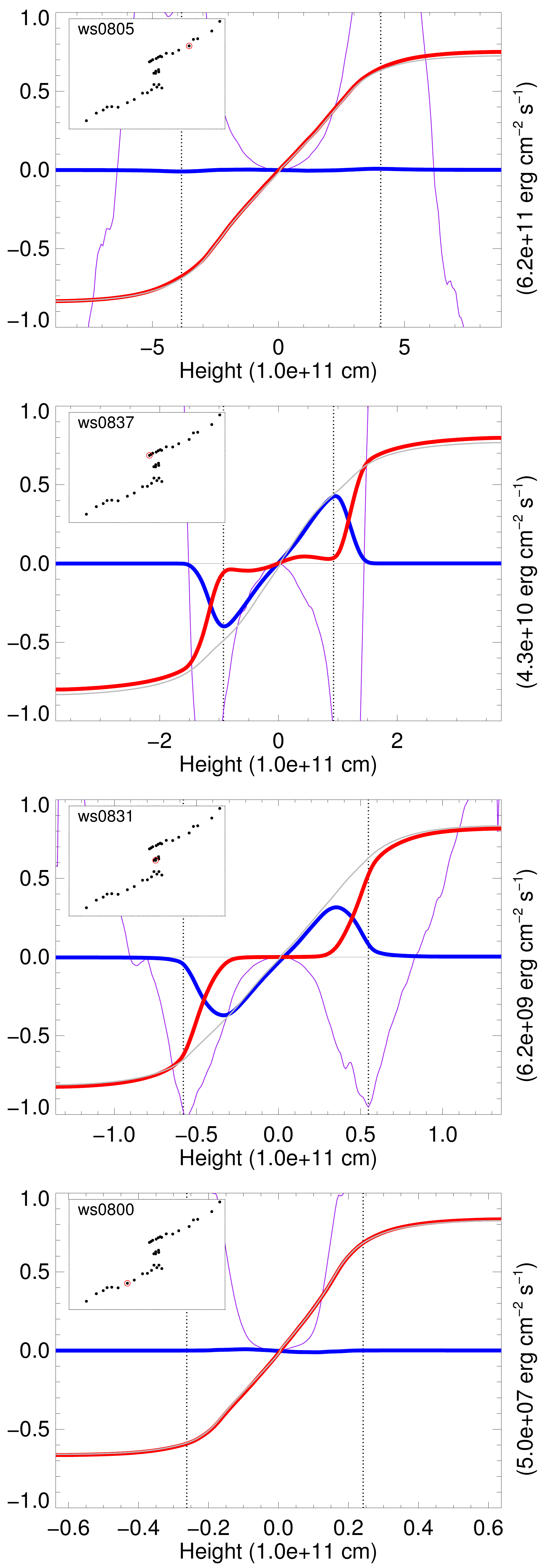}
  \caption{
    Time and horizontally-averaged vertical profiles of radiative heat flux $\bar{F}^-_\text{rad}$ (red),
    advective heat flux $\bar{F}^-_\text{adv}$ (blue), and cumulative heating rate $\bar{F}^+_\text{heat}$ (gray),
    for selected solutions.
    In the {\color{\minor}inset} in each panel,
    the position of the solution on the $\bar{\Sigma}$--$\bar{T}_\text{eff}$ plane (Figure \ref{fig:s-curve}) is indicated by a red circle.
    In each panel, the fluxes are normalized by the value shown on the right axis. 
    The purple curve shows the hydrodynamic Brunt-V\"{a}is\"{a}l\"{a} frequency squared divided by the angular velocity squared, $N^2/\Omega^2$.
    The plasma beta is larger than unity at the heights between the two vertical dotted lines.
  \label{fig:energyflux}}
\end{figure}

\begin{figure}
  \centering
  \subfloat{\includegraphics[width=8.5cm]{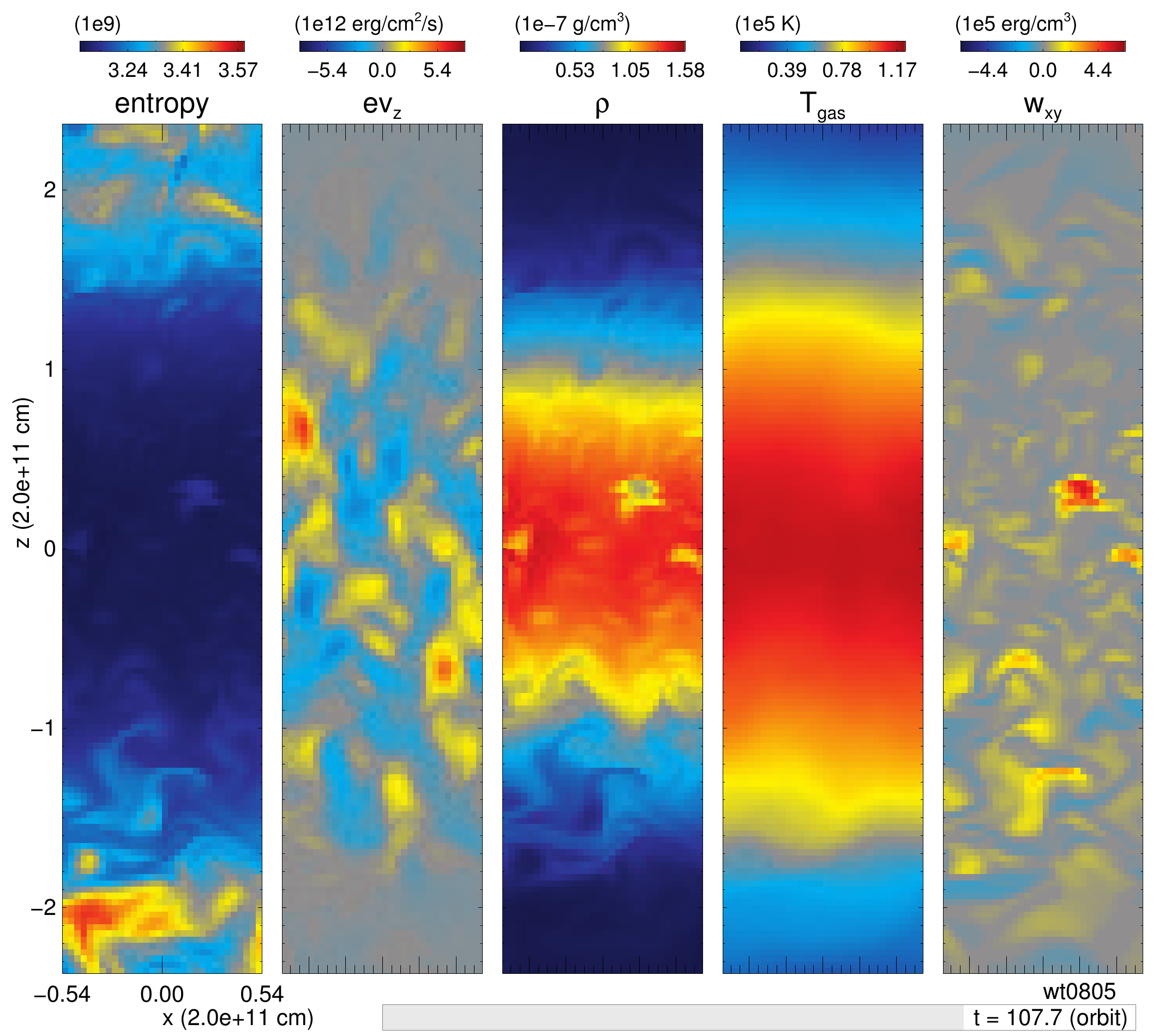}}\\ 
  \subfloat{\includegraphics[width=8.5cm]{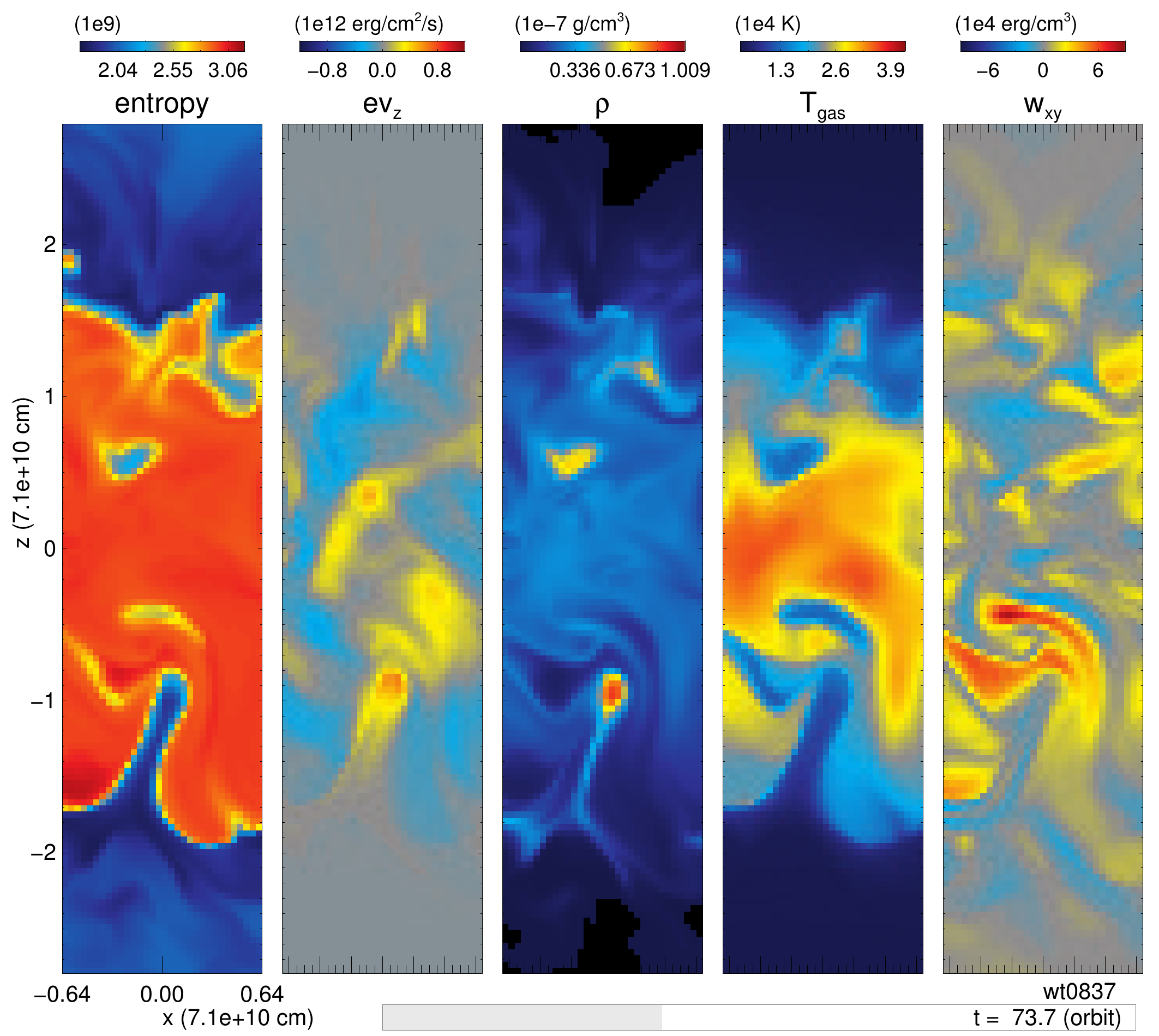}}
  \caption{
    Snapshots of various quantities on an $x$-$z$ plane ($y = 0$) for a radiative solution (upper, $t=107.7$ orbits in ws0805) and a convective solution
    (lower, $t=73.7$ orbits in ws0837).
    From the left to the right, specific entropy, vertical advective heat flux $ev_z$, density $\rho$, gas temperature $T$,
    and the total stress $w_{xy}$ are shown.
    The $x$ and $z$ axes are normalized by the time-averaged pressure scale height $\bar{h}_\text{p}$ (see Table \ref{table}). 
    Note that images here do not include the entire vertical extent of the box, but instead are limited to the midplane regions.
    Movies of snapshots of these two simulations are available online.
  }
  \label{fig:slice}
\end{figure}

\begin{figure}
  \centering
  \includegraphics[width=6.5cm]{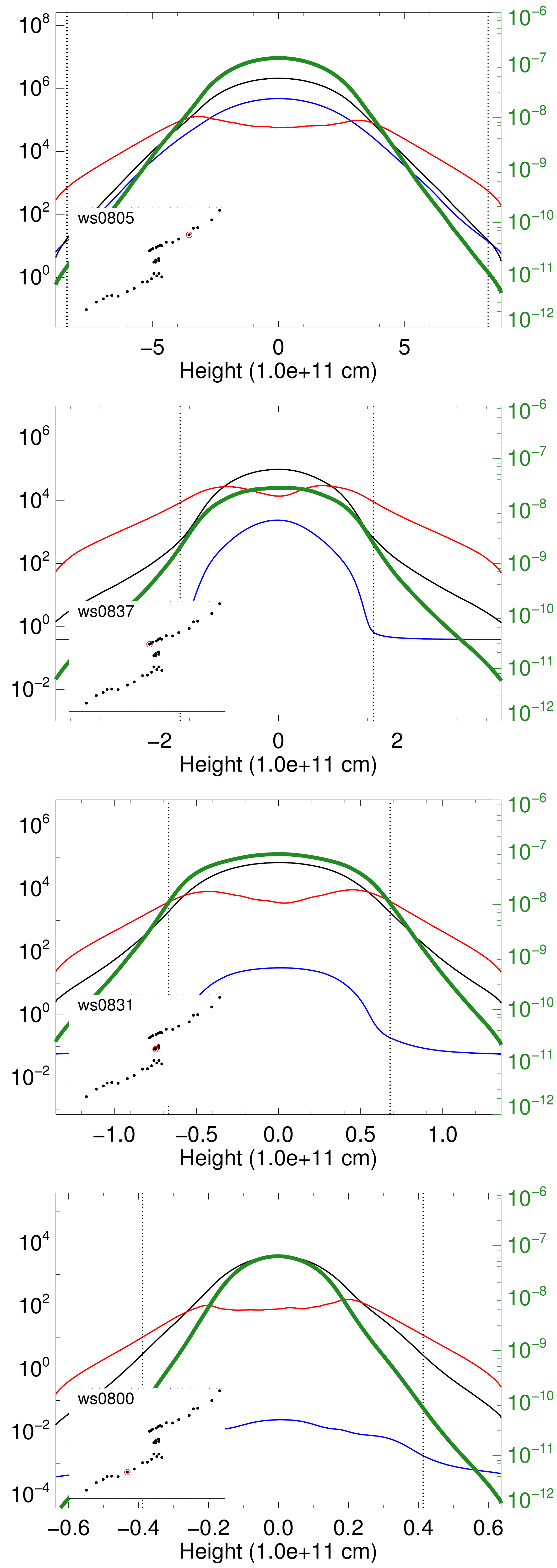}
  \caption{
    Time and horizontally-averaged vertical profiles of density (green), gas pressure (black), magnetic pressure (red), and radiation pressure (blue)
    for the selected solutions in Figure \ref{fig:energyflux}. The unit of pressures is ergcm$^{-3}$ and the axis for density (gcm$^{-3}$) is on the right.
    The vertical dotted lines denote the heights where the Rosseland-mean optical depth from the top/bottom boundary is unity.
    Other notations are the same as in Figure \ref{fig:energyflux}.}
  \label{fig:density}
\end{figure}

\begin{figure}
  \centering
  \includegraphics[width=6.5cm]{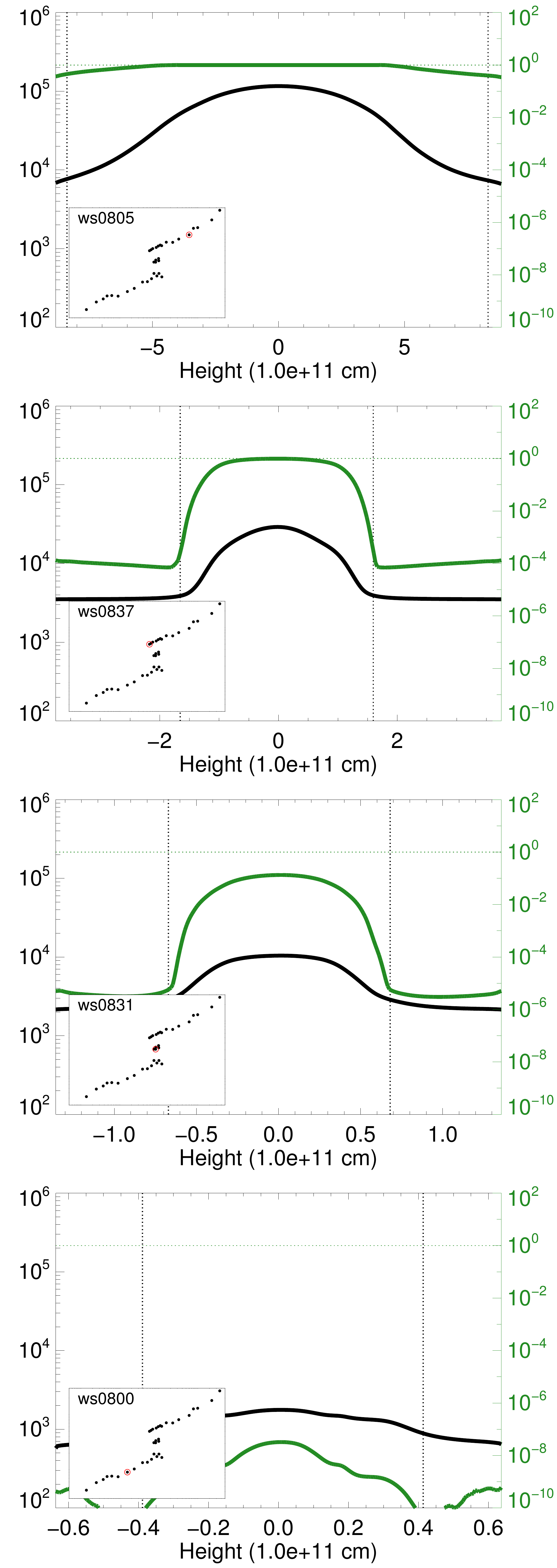}
  \caption{
    Time and horizontally-averaged vertical profiles of gas temperature (K) (black) and
    ionization fraction (green) for the selected solutions in Figure \ref{fig:energyflux}. 
    The axis for the ionization fraction is on the right.
    Other notations are the same as in Figure \ref{fig:density}.}
  \label{fig:temperature}
\end{figure}

\begin{figure}
  \centering
  \includegraphics[width=7cm]{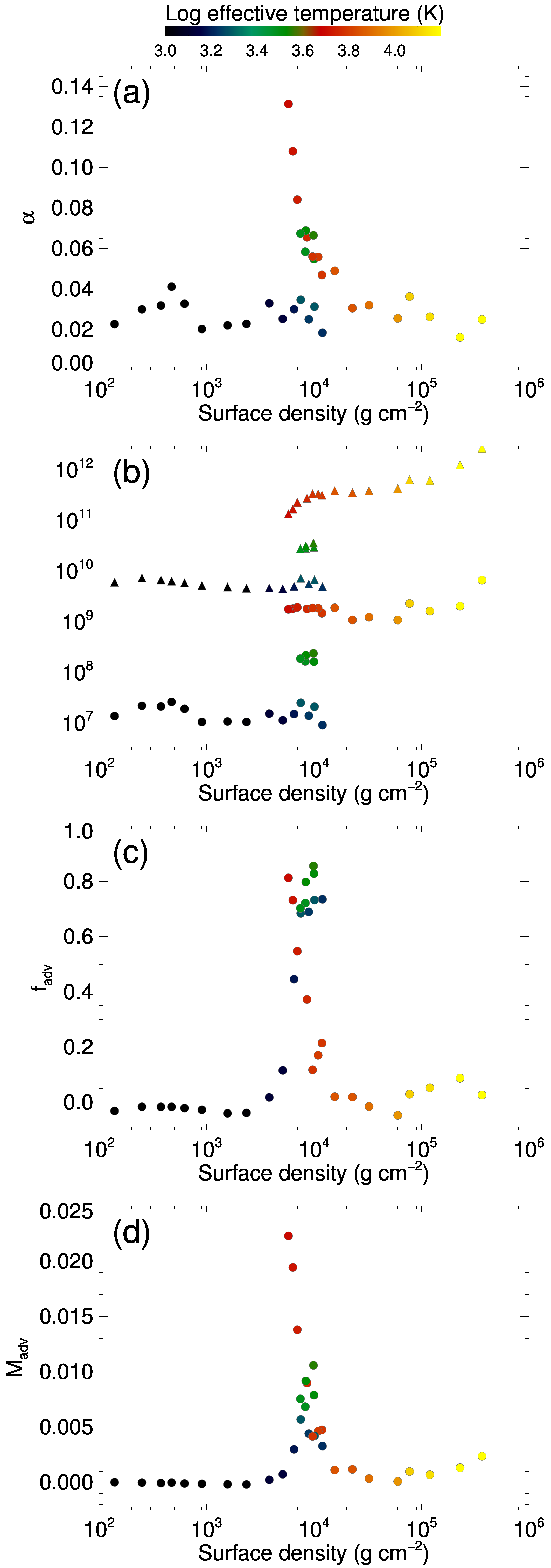}
  \caption{
    Time-averaged quantities as a function of surface density $\bar{\Sigma}$:
    (a) $\bar{\alpha}$,
    (b) vertically-integrated stress $\bar{W}_{xy} (\times0.1)$ (circles) and vertically-integrated thermal
    pressure $\bar{P}_\text{thermal}$ (triangles), each divided by $\bar{\Sigma}^{4/3}$,
    (c) advective fraction in the energy transport $\bar{f}_\text{adv}$, and
    (d) Mach number of the convective motion $\bar{M}_\text{adv}$.
    Colors represent the time-averaged effective temperature $\bar{T}_\text{eff}$; the reddish and yellowish symbols correspond to the upper branch solutions
    while dark-greenish and dark-bluish symbols correspond to the middle and the lower branch solutions, respectively.}
  \label{fig:correlation}
\end{figure}

\begin{figure}
  \centering
  \includegraphics[width=7cm]{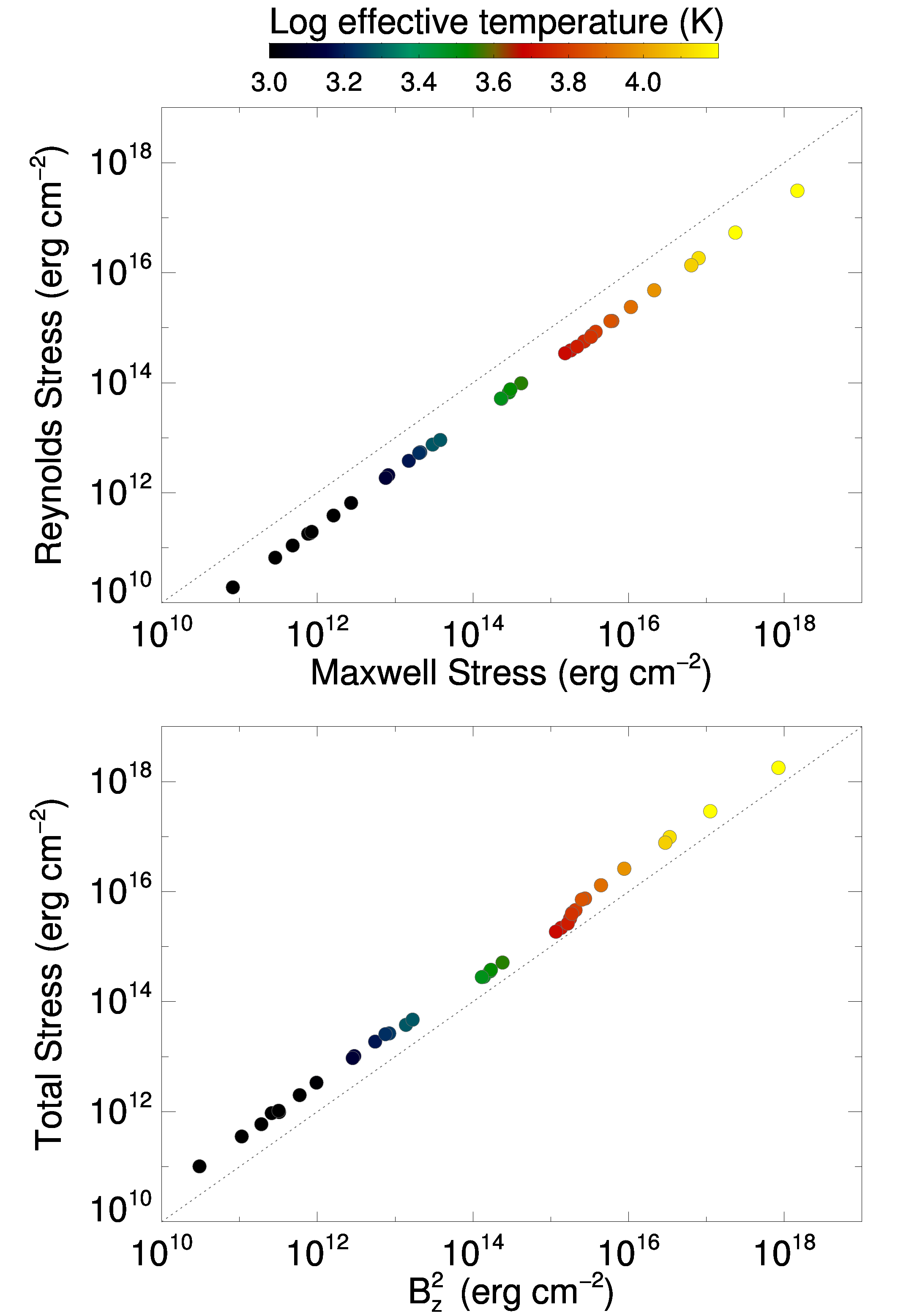}
  \caption{
    Time-averaged vertically-integrated Reynolds stress vs. Maxwell stress (upper) and
    vertically-integrated total stress $\bar{W}_{xy}$ vs. $\int\left< B_z^2\right> dz$ (lower).
    The dotted line indicates the {\color{\minor}equality}.
    Other notations are the same as in Figure \ref{fig:correlation}.}
  \label{fig:stressbz2}
\end{figure}

\begin{figure}
  \centering
  \includegraphics[width=7cm]{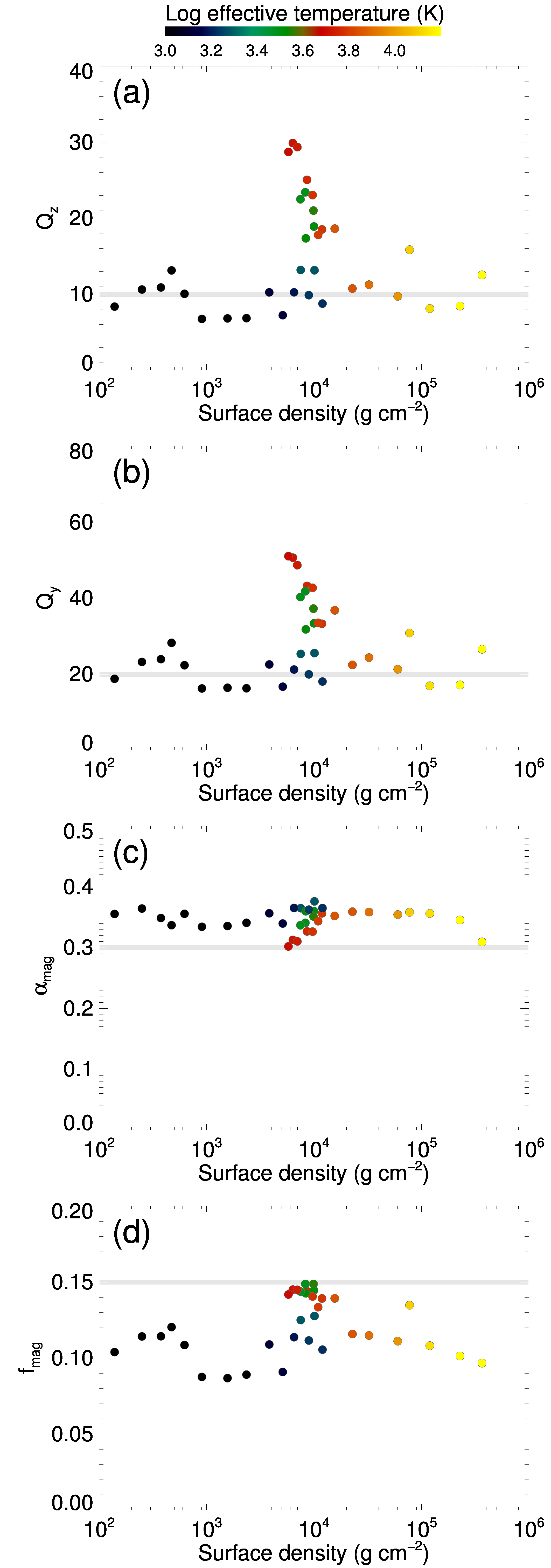}
  \caption{\color{\major}
    Convergence metrics (a) $Q_z$, (b) $Q_y$, (c) $\alpha_\text{mag}$, and (d) $f_\text{mag}$ (see text for the definitions) as a function of surface density $\bar{\Sigma}$.
    The gray horizontal line in panels (a) and (b) indicates a criterion for the numerical convergence while that in panels (c) and (d) indicates a criterion for well-developed MRI turbulence, suggested by \citet{Hawley11}.
    Other notations are the same as in Figure \ref{fig:correlation}.}
  \label{fig:assess}
\end{figure}

\begin{figure}
  \centering
  \includegraphics[width=7cm]{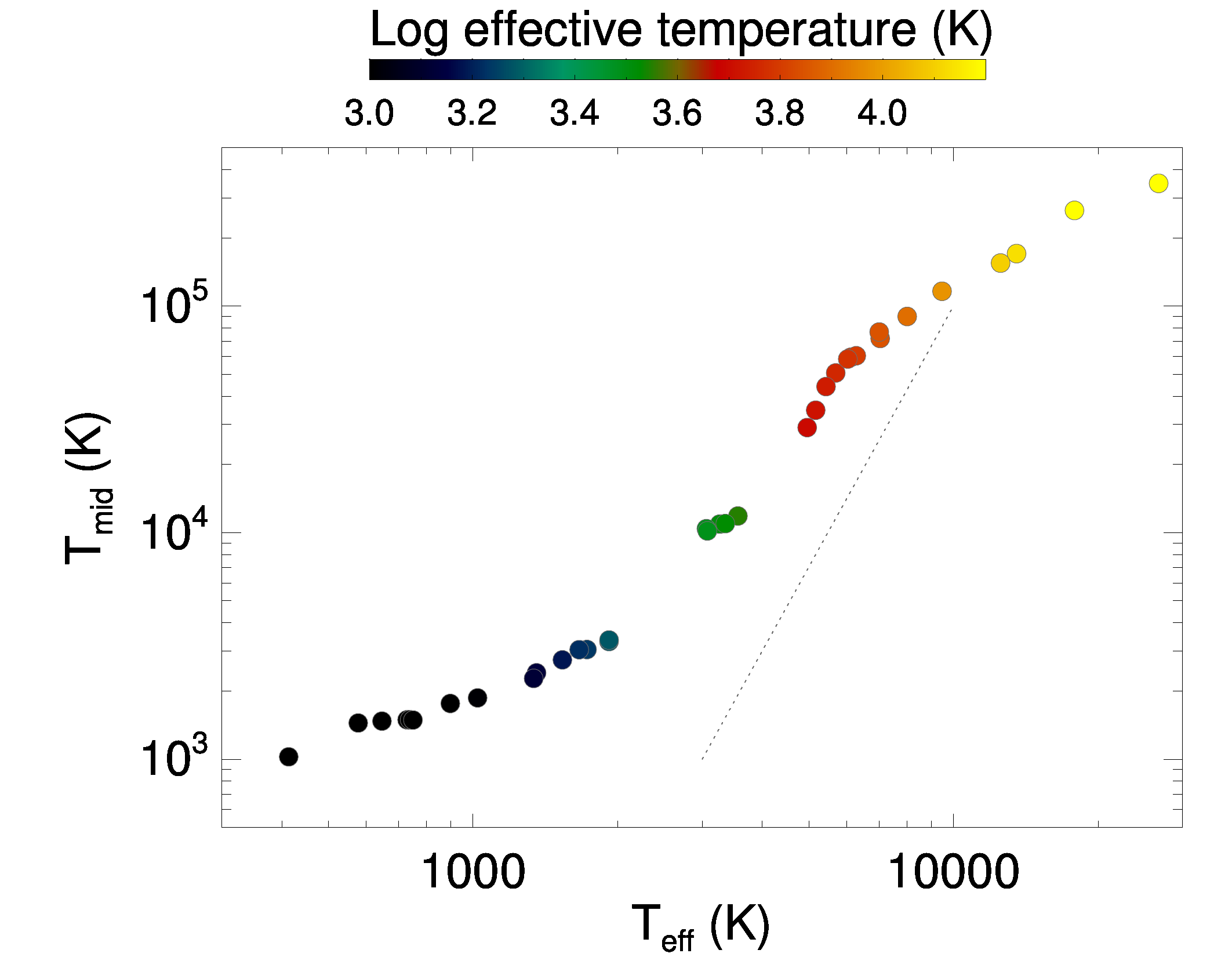}
  \caption{
    Time-averaged midplane temperature $\bar{T}_\text{mid}\equiv \left<T\right>(z=0)$ vs. effective temperature $\bar{T}_\text{eff}$.
    The dotted line indicates the critical slope, $T_\text{mid} \propto T_\text{eff}^4$. 
    Other notations are the same as in Figure \ref{fig:correlation}.}
  \label{fig:tmidteff}
\end{figure}

\begin{figure}
  \centering
  \includegraphics[width=7cm]{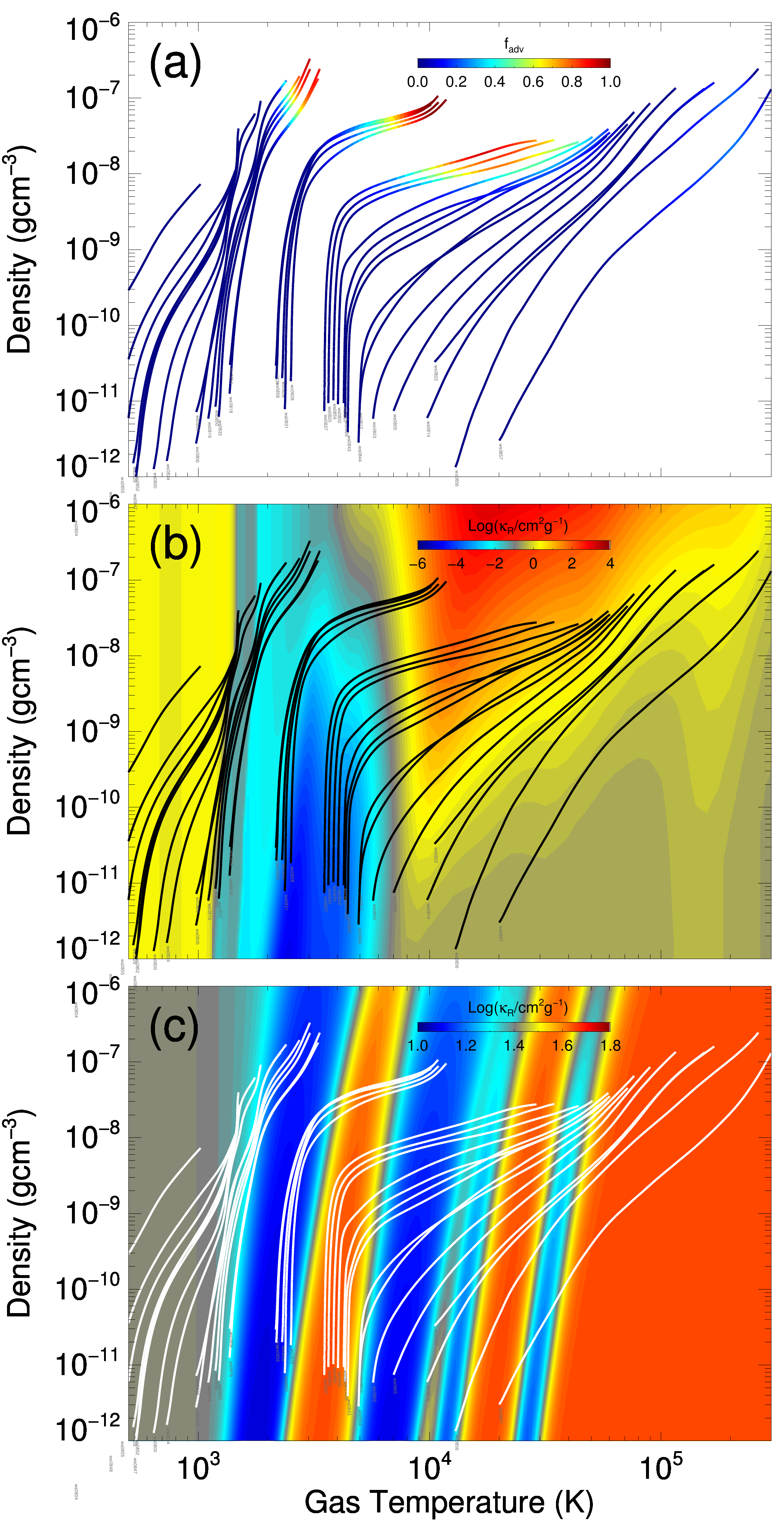}
  \caption{
    Time and horizontally-averaged vertical structure $(\left<{T}\right>(z),\left<{\rho}\right>(z))$ 
    plotted on the $T$--$\rho$ plane for all runs that reached a thermal equilibrium.
    The midplane and the photosphere correspond to the top and the bottom of each curve, respectively.
    Curves are colored by the advective fraction $\bar{f}_\text{adv}$ in panel (a) while
    the background color indicates the Rosseland-mean opacity $\kappa^\text{R}(T,\rho)$ cm$^2$g$^{-1}$ in panel (b) and
    the first generalized adiabatic exponent $\Gamma_1(T,\rho)$ in panel (c).}
  \label{fig:denstemp}
\end{figure}

\begin{figure}
  \centering
  \includegraphics[width=5.35cm]{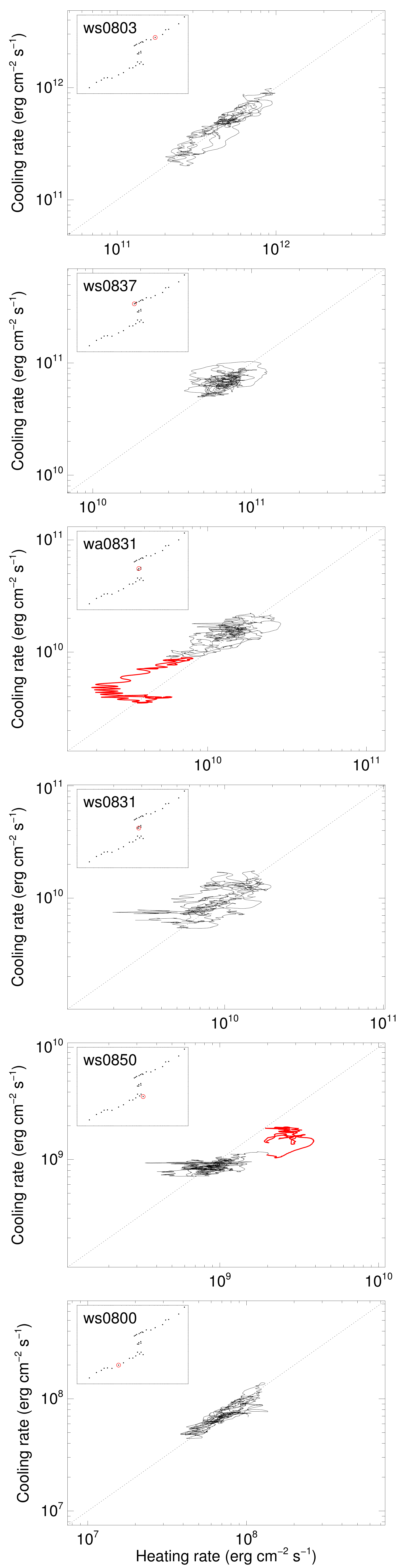}
  \caption{
    Time trajectory of $(\tilde{Q}^+(t),\tilde{Q}^-(t))$ during the time period $\Delta$ (see equation \ref{eq:time_averaging}), for selected solutions.
    The portion corresponding
    to the last 15 orbits is indicated by color red for wa0831 and ws0850, which finally showed runaway cooling and heating, respectively.
    The displayed ranges of $\tilde{Q}^+$ and $\tilde{Q}^-$ is always two orders of magnitude.
    Other notations are the same as in Figure \ref{fig:energyflux}.}
  \label{fig:stability}
\end{figure}

\begin{figure}
  \centering
  \includegraphics[width=7cm]{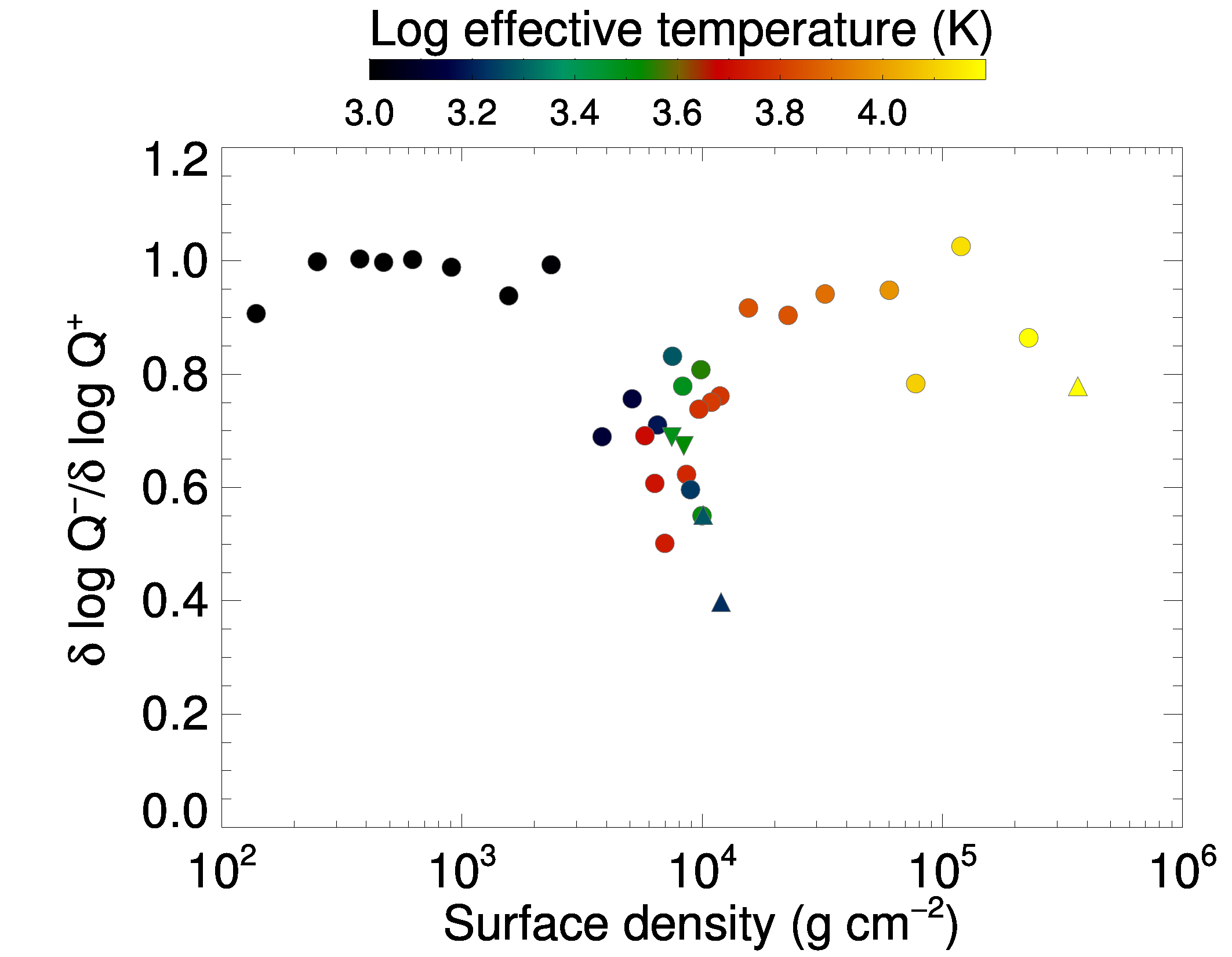}
  \caption{
    The ratio of standard deviation of logarithmic cooling rate $\log \tilde{Q}^-(t)$ to
    that of logarithmic heating rate $\log \tilde{Q}^+(t)$ (see Figure \ref{fig:stability}).
    The upward (downward) triangles indicate the solutions that showed runaway heating (cooling) in the end, respectively.
    Other notations are the same as in Figure \ref{fig:correlation}.}
  \label{fig:stability2}
\end{figure}
\clearpage

\appendix
\section{Radiation Damping As a Heat Source}\label{sec:raddamping}
\begin{figure}
  \centering
  \includegraphics[width=6.5cm]{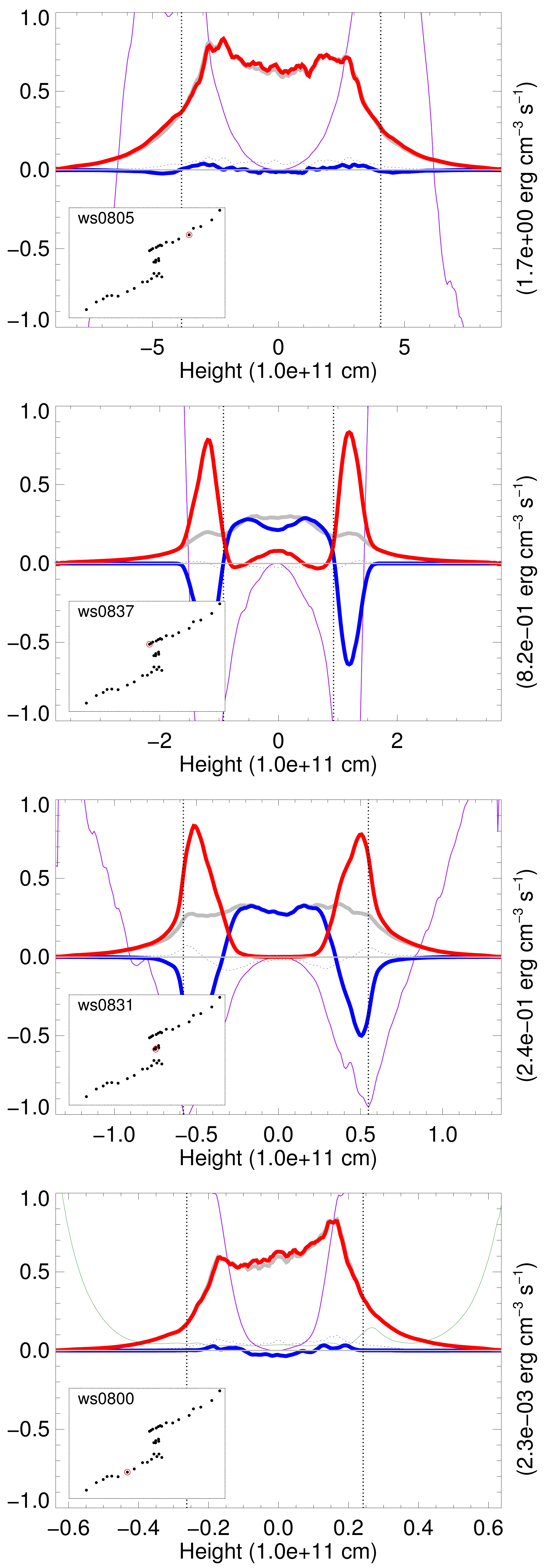}
  \caption{
    Time and horizontally-averaged vertical profiles of radiative cooling rate $\bar{q}^-_\text{rad}$ (red),
    advective cooling rate $\bar{q}^-_\text{adv}$ (blue), the compressional heating rate (dotted) $\bar{q}^+_\text{comp}$,
    and the total heating rate $\bar{q}^+_\text{total}$ (gray),
    for the selected solutions in Figure \ref{fig:energyflux}, 
    normalized by the value shown on the right axis in each panel. 
    The green curve in the panel of ws0800 shows the {\color{\minor}acoustic} radiation damping rate $\bar{q}^+_\text{rad}$ (\ref{eq:c4}){\color{\minor},
      which may not be applicable in high
      altitudes where the slow diffusion limit and the ignorance of
      magnetic field are invalid.}
    Other notations are the same as in Figure \ref{fig:energyflux}.}
  \label{fig:cooling}
\end{figure}

Figure \ref{fig:cooling} shows time and horizontally-averaged vertical profiles of cooling and heating rates (see equation \ref{eq:balance}),
\begin{align}
  &\bar{q}_\text{total}^+(z) \equiv \left< q_\text{diss}^+\right> - \left<\left(\nabla\cdot\bm{v}\right)p - \nabla\bm{v}:\mathsf{P}\right>, \\
  &\bar{q}_\text{comp}^+(z) \equiv - \left<\left(\nabla\cdot\bm{v}\right)p - \nabla\bm{v}:\mathsf{P}\right>, \\
  &\bar{q}_\text{rad}^-(z) \equiv \left<\dfrac{dF_z}{dz}\right>, \\
  &\bar{q}_\text{adv}^-(z) \equiv \left<\dfrac{d}{dz}\left((e + E)v_z\right)\right>,
\end{align}
which are differential versions of those shown in Figure \ref{fig:energyflux}.
Therefore, here we confirm again thermal equilibrium, where the total cooling rate and the total heating rate match.
{\color{\minor}We note that most of the dissipation occurs inside the Rosseland-mean photospheres in all cases, which justifies
  the first equality in equation (\ref{eq:thermal_equilibrium}).}
The compressional heating rate $q_\text{comp}^+$ should be zero when vertically-integrated if the fluid motion is adiabatic.
However, it can be positive when the fluid motion is damped by radiative diffusion, which is the case here.
For example, in ws0800, $\int \bar{q}^+_\text{comp}dz \sim 0.096 \int \bar{q}^{+}_\text{total} dz$.
{\color{\minor}
  That is, about 10 \% of the total heating comes from the virtual dissipation associated with compressional heating.
  We also computed the vertical component of $\int \bar{q}^+_\text{comp} dz$ to find
  $\int \left<p\right>(d\left<v_z\right>/dz)dz \sim -0.019 \int \bar{q}^{+}_\text{total} dz$. This means that
  the compressional heating comes mostly from horizontal compressions ($= \int \bar{q}^+_\text{comp}dz - \int \left<p\right>(d\left<v_z\right>/dz)dz \sim 0.115 \int \bar{q}^{+}_\text{total} dz$),
  presumably due to MRI turbulence and the spiral acoustic waves.}
{\color{\minor}Below we evaluate the radiation damping rate $q_\text{rad}^+$ of an acoustic wave
  (that represent the fluid motion) to show $q_\text{rad}^+ \sim
  q_\text{comp}^+$, which will confirm that the non-zero (positive) compressional heating
  comes from the radiation damping.}

First, we start from the dispersion relation given in equation (1) in \citet{Blaes07} to get the exponential damping rate of the amplitude of a linear, plane acoustic wave:
\begin{align}
  \omega^2 &= \frac{\omega(e+4E)c_\text{t}^2 + (ick^2/3\kappa_F\rho)4Ec_\text{i}^2}{\omega(e+4E) + (ick^2/3\kappa_F\rho)4E}k^2 \label{eq:a1}\\
  &\cong \left(c_\text{t}^2 + \frac{(ick^2/3\kappa_F\rho)4Ec_\text{i}^2}{\omega(e+4E)}\right)\left(1-\frac{(ick^2/3\kappa_F\rho)4E}{\omega(e+E)}\right)k^2  \label{eq:a2}\\
  &\cong c_\text{t}^2k^2\left(1 - \frac{(ck^2/3\kappa_F\rho)4E}{\omega(e+4E)}\left(1-\frac{c_\text{i}^2}{c_\text{t}^2}\right)i\right)  \label{eq:a3}\\
  &\cong c_\text{t}^2k^2\left(1 - \frac{(ck^2/3\kappa_F\rho)4E}{c_\text{t}k(e+4E)}\left(1-\frac{c_\text{i}^2}{c_\text{t}^2}\right)i\right), \label{eq:a4}
\end{align}
where $\omega$ and $k$ are the frequency and the wave number of the acoustic wave, respectively, $\kappa_F$ the flux-mean opacity, $c_\text{i}$ the isothermal sound speed,
and $c_\text{t}$ the total sound speed.
From (\ref{eq:a1}) to (\ref{eq:a2}), we take the slow diffusion limit,
where the first term dominates the second term both in the numerator and the denominator in the right hand side of (\ref{eq:a1}).
Then, we take $\omega^2 = c_\text{t}^2k^2$ (the undamped acoustic wave) to lowest order,
and substitute it in the right hand side of (\ref{eq:a3}) to get to (\ref{eq:a4}),
where $\omega$ is expressed as a function of $k$.
Then, the damping rate of the amplitude $\Gamma$ is obtained as the imaginary part of $\omega$:
\begin{align}
  \Gamma &\cong \frac12\frac{(ck^2/3\kappa_F\rho)4E}{(e+4E)}\left(1-\frac{c_\text{i}^2}{c_\text{t}^2}\right) \label{eq:b1}\\
  &\cong\begin{cases}
  \dfrac{2ck^2}{3\kappa_F\rho}\dfrac{e}{E}\left(1-\dfrac{1}{\Gamma_1}\right) & (e\gg E)\\
  \dfrac{ck^2}{6\kappa_F\rho} & (e\ll E), \label{eq:b2} 
  \end{cases}
\end{align}
where the factor $1/2$ in the right hand side of (\ref{eq:b1}) comes from Taylor expansion of square root of (\ref{eq:a4}).
Hereafter, we assume that $e \ll E$, and the total pressure is replaced with gas pressure.
Then, as derived in equation (24) in \citet{Blaes11}, the acoustic radiative damping rate can be evaluated as
\begin{align}
  q^+_\text{rad} &\cong \frac{(\delta p_\text{max})^2}{\rho c_\text{t}^2}\Gamma \label{eq:c1}\\
  &\cong \left(\frac{\delta p_\text{max}}{p}\right)^2\frac{p}{\Gamma_1}\Gamma \label{eq:c2}\\
  &\cong \frac{1}{\Gamma_1}\left(1-\frac{1}{\Gamma_1}\right)\frac{2ckp}{3\kappa_\text{R}\rho/k}\frac{e}{E}\left(\frac{\delta p_\text{max}}{p}\right)^2, \label{eq:c3}
\end{align}
where $\delta p_\text{max}$ is the pressure amplitude of the acoustic wave.
From (\ref{eq:c2}) to (\ref{eq:c3}), (\ref{eq:b2}) was substituted with the approximation $\kappa_F \approx \kappa_\text{R}$.

\begin{figure}
  \centering
  \includegraphics[width=4cm]{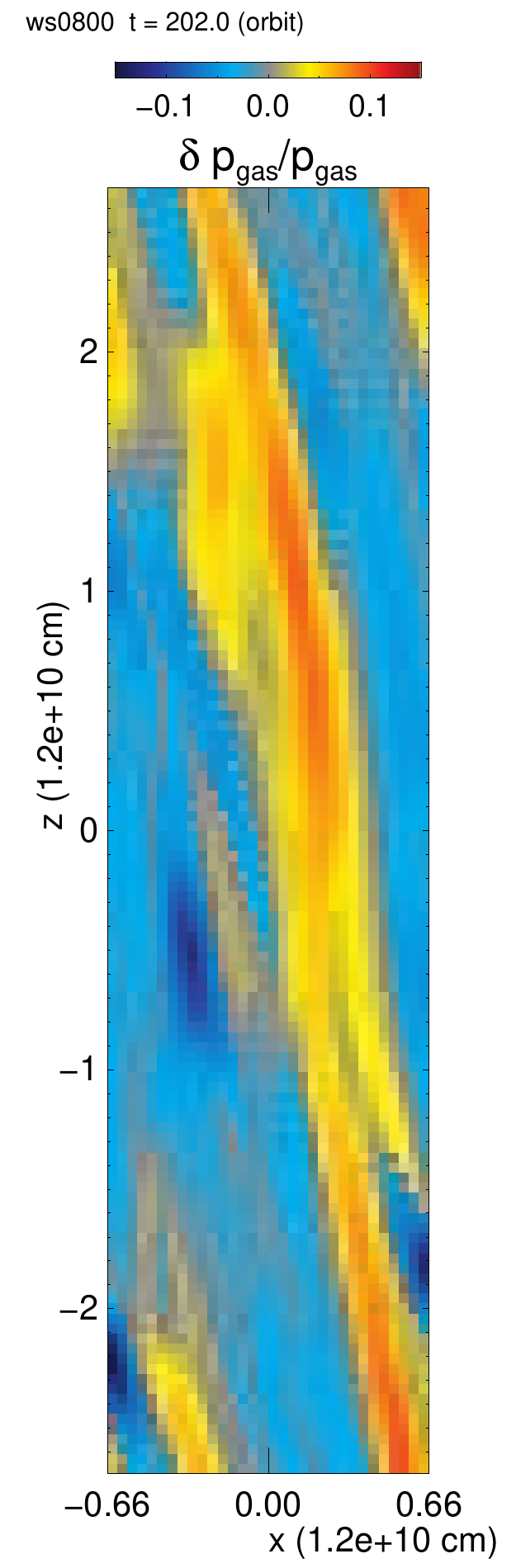}
  \caption{
    Fluctuation of gas pressure relative to its horizontal average on the midplane
    at $t=202.0$ orbits in the lower branch solution, ws0800. Axes are normalized by the time-averaged pressure scale height $\bar{h}_\text{p}$.}
  \label{fig:pres_slice}
\end{figure}

Using simulation results, we can compute the time-averaged version of (\ref{eq:c3}) as 
\begin{align}
  \bar{q}_\text{rad}^+(z) \equiv \frac{1}{\left<\Gamma_1\right>}\left(1-\frac{1}{\left<\Gamma_1\right>}\right)\frac{2ck\left< p\right>}{3\left<\kappa_\text{R}\rho\right>/k}\frac{\left< e\right>}{\left< E\right>}\left(\frac{\delta p_\text{max}}{p}\right)^2,\label{eq:c4}
\end{align}
where the wave number $k$ and the pressure fluctuation $\delta p_\text{max}/p$ may be evaluated from a snapshot of a simulation.
Here, we take ws0800 as an example, and in that case, they are evaluated from Figure \ref{fig:pres_slice} as 
\begin{align}
  &k \cong \frac{2\pi}{\text{(box width in $x$)}} = \frac{2\pi}{1.6\times10^{10}(\text{cm})} = 3.9\times10^{-10}\text{(cm$^{-1}$)},\\
  &\frac{\delta p^\text{max}}{p} \cong 0.13.
\end{align}
As shown in the panel of ws0800 in Figure \ref{fig:cooling}, 
the resultant radiation damping rate $\bar{q}_\text{rad}^+(z)$ shows a good agreement with
the compressional heating rate $\bar{q}_\text{comp}^+(z)$ for the heights where $\bar{q}_\text{comp}^+(z)$ is consistently positive.
Therefore, the consistent positive compressional heating is regarded as the acoustic radiation damping.

\clearpage


\label{lastpage}

\end{document}